\documentclass[conference,compsoc]{IEEEtran}
\usepackage{cite}
\usepackage{amsmath,amssymb,amsfonts}
\usepackage{algorithmic}
\usepackage{graphicx}
\usepackage{textcomp}
\usepackage{xcolor}
\usepackage[hyphens]{url}

\usepackage[utf8]{inputenc}
\usepackage[T1]{fontenc}
\usepackage[american]{babel}
\usepackage{booktabs, tabularx, multirow}
\usepackage{tablefootnote}
\usepackage[nohyperlinks,nolist]{acronym}
\usepackage[operators,sets,primitives,keys,notions,advantage,adversary,logic,ff,asymptotics,probability]{cryptocode}
\usepackage{siunitx}
\usepackage[amsmath,thmmarks]{ntheorem}
\usepackage{wasysym}
\usepackage{enumitem}
\usepackage[linkcolor=black,colorlinks=true,citecolor=black,filecolor=black,urlcolor=black]{hyperref}
\usepackage[hyphenbreaks]{breakurl}
\usepackage[nameinlink]{cleveref}
\usepackage{balance}

\crefname{proof}{proof}{proofs}
\crefname{attack}{attack}{attacks}
\Crefname{attack}{Attack}{Attacks}

\newcommand{\fakeparagraph}[1]{\vskip 5pt\noindent\textbf{#1.}}

\begin{document}
\title{
\vspace{-1.75em}
Verifiable Fully Homomorphic Encryption \vspace{-0.75em}}
\author{%
\IEEEauthorblockN{Alexander Viand\textsuperscript{*}, Christian Knabenhans\textsuperscript{*}, Anwar Hithnawi} \\
\IEEEauthorblockA{\textit{ETH Zurich}}
 }

\maketitle
\begingroup\renewcommand\thefootnote{*}
\footnotetext{Equal contribution}
\endgroup

\thispagestyle{plain}
\pagestyle{plain}

\begin{abstract}
\acf{FHE} is seeing increasing real-world deployment to protect data in use by allowing computation over encrypted data.
However, the same malleability that enables homomorphic computations also raises \emph{integrity} issues, which have so far been mostly overlooked.
While FHE's lack of integrity has obvious implications for correctness,
it also has severe implications for confidentiality:
a malicious server can leverage the lack of integrity to carry out interactive key-recovery attacks.
As a result, virtually all FHE schemes and applications assume an honest-but-curious server who does not deviate from the protocol.
In practice, however, this assumption is insufficient for a wide range of deployment scenarios.
While there has been work that aims to address this gap, these have remained isolated efforts considering only aspects of the overall problem and fail to fully address the needs and characteristics of modern FHE schemes and applications.
In this paper,
we analyze existing FHE integrity approaches, present attacks that exploit gaps in prior work, and propose a new notion for maliciously-secure verifiable FHE.
We then instantiate this new notion with a range of techniques, analyzing them and evaluating their performance in a range of different settings.
We highlight their potential but also show where future work on tailored integrity solutions for FHE is still required.
\end{abstract}

\numberwithin{equation}{section}

\theoremstyle{break}
\theoremheaderfont{\normalfont\bfseries}
\newtheorem{theorem}{Theorem}[section]
\newtheorem{example}[theorem]{Example}
\newtheorem{definition}[theorem]{Definition}

\theoremstyle{nonumberplain}
\theorembodyfont{\normalfont}
\theoremsymbol{}%
\newtheorem{proof}{Proof}

\newcommand{\work}{paper}

\definecolor{ETHBlue}{RGB}{33,92,175}	%
\definecolor{ETHGreen}{RGB}{98,115,19}		%
\definecolor{ETHPurple}{RGB}{163,7,116}	%
\definecolor{ETHGray}{RGB}{111,111,111}	%
\definecolor{ETHRed}{RGB}{183,53,45}	%
\definecolor{ETHPetrol}{RGB}{0,120,148}	%
\definecolor{ETHBronze}{RGB}{142,103,19}	%

\colorlet{ETHdarkblue}{ETHBlue}
\colorlet{ETHdarkgreen}{ETHGreen}
\colorlet{ETHpink}{ETHPurple}
\colorlet{ETHgray}{ETHGray}
\colorlet{ETHred}{ETHRed}
\colorlet{ETHgreenblue}{ETHPetrol}
\colorlet{ETHbrown}{ETHBronze}

\newcommand{\E}[1]{\ensuremath{\pcalgostyle{E}\left(#1\right)}}
\newcommand{\Einvnostar}[1]{\ensuremath{\pcalgostyle{E}^{-1}\left(#1\right)}}
\newcommand{\Einvstar}[1]{\ensuremath{\pcalgostyle{E}^{-1}(#1)}}
\NewDocumentCommand\Einv{sm}{%
  \IfBooleanTF#1{\Einvstar{#2}}{\Einvnostar{#2}}
}

\newcommand{\setup}{\ensuremath{\pcalgostyle{Setup}}}
\newcommand{\prove}{\ensuremath{\pcalgostyle{Prove}}}
\renewcommand{\verify}{\ensuremath{\pcalgostyle{Verify}}}

\newcommand{\recompute}{\ensuremath{\pcalgostyle{Recompute}}}
\newcommand{\trans}{\ensuremath{\pckeystyle{qt}}}
\newcommand{\extract}{\ensuremath{\pcalgostyle{Extract}}}
\newcommand{\Gen}{\ensuremath{\pcalgostyle{Gen}}}

\newcommand{\probgen}{\ensuremath{\pcalgostyle{ProbGen}}}
\newcommand{\pubprobgen}{\ensuremath{\pcalgostyle{PubProbGen}}}
\newcommand{\compute}{\ensuremath{\pcalgostyle{Compute}}}
\newcommand{\decode}{\ensuremath{\pcalgostyle{Decode}}}
\newcommand{\valid}{\ensuremath{\pcalgostyle{Valid}}}
\newcommand{\pubkgen}{\ensuremath{\pcalgostyle{PubKGen}}}
\newcommand{\pubverify}{\ensuremath{\pcalgostyle{PubVerify}}}

\newcommand{\comscheme}{\ensuremath{\pcalgostyle{Com}}}
\newcommand{\commit}{\ensuremath{\pcalgostyle{Commit}}}
\newcommand{\comver}{\ensuremath{\pcalgostyle{ComVer}}}
\newcommand{\open}{\ensuremath{\pcalgostyle{Open}}}
\newcommand{\ck}{\ensuremath{\pckeystyle{ck}}}
\newcommand{\com}{\ensuremath{\pcalgostyle{com}}}

\newcommand{\estimate}{\ensuremath{\pcalgostyle{Estimate}}}

\newcommand{\pcas}{\highlightkeyword{~as}}

\newcommand{\cheq}{\overset{?}{=}}

\newcommand{\crs}{\key[crs]}
\newcommand{\td}{\key[td]}
\renewcommand{\state}{\key[st]}
\renewcommand{\secpar}{\ensuremath{\lambda}}
\renewcommand{\secparam}{\ensuremath{1^\secpar}}

\newcommand{\Iio}{\ensuremath{I_\te2xt{io}}}
\newcommand{\Imid}{\ensuremath{I_\text{mid}}}
\newcommand{\Iin}{\ensuremath{I_\text{in}}}
\newcommand{\Iout}{\ensuremath{I_\text{out}}}

\newcommand{\Pin}{\ensuremath{M_\text{in}}}
\newcommand{\Pout}{\ensuremath{M_\text{out}}}
\newcommand{\Rin}{\ensuremath{R_\text{in}}}
\newcommand{\Oin}{\ensuremath{O_\text{in}}}
\newcommand{\Comin}{\ensuremath{\pcalgostyle{Com}_\text{in}}}

\newcommand{\Cio}{\ensuremath{C_\text{io}}}
\newcommand{\Cin}{\ensuremath{C_\text{in}}}
\newcommand{\Cout}{\ensuremath{C_\text{out}}}
\newcommand{\cout}{\ensuremath{c_\text{out}}}
\newcommand{\Cmid}{\ensuremath{C_\text{mid}}}

\newcommand{\Piin}{\ensuremath{\Pi_\text{in}}}
\newcommand{\piout}{\ensuremath{\pi_\text{out}}}

\newcommand{\iin}{\ensuremath{i_\text{in}}}
\newcommand{\iout}{\ensuremath{i_\text{out}}}
\newcommand{\Sin}{\ensuremath{S_\text{in}}}
\newcommand{\Sout}{\ensuremath{S_\text{out}}}

\newcommand{\vmid}{\ensuremath{v_\text{mid}}}
\newcommand{\wmid}{\ensuremath{w_\text{mid}}}
\newcommand{\ymid}{\ensuremath{y_\text{mid}}}
\newcommand{\vio}{\ensuremath{v_\text{io}}}
\newcommand{\wio}{\ensuremath{w_\text{io}}}
\newcommand{\yio}{\ensuremath{y_\text{io}}}

\newcommand{\pt}{\ensuremath{\mathbf{m}}}
\newcommand{\ct}{\ensuremath{\mathbf{ct}}}
\newcommand{\ctout}{\ensuremath{\ct_\text{out}}}

\newcommand{\modb}[2]{\ensuremath{\left[#1\right]_{#2}}}
\newcommand{\round}[1]{\ensuremath{\left\lfloor#1\right\rceil}}
\newcommand{\tup}[1]{\ensuremath{\left(#1\right)}}
\newcommand{\setst}[2]{\ensuremath{\set{#1\,\left|\,#2\vphantom{#1}\right.}}}

\newcommand{\ptstyle}[1]{\ensuremath{\mathbf{#1}}}
\newcommand{\ctstyle}[1]{\ensuremath{\mathbf{#1}}}

\newcommand{\op}[1]{\ensuremath{\texttt{#1}}}
\newcommand{\rel}{\ensuremath{\mathcal{R}}}
\newcommand{\lang}{\ensuremath{\mathcal{L}}}

\newcommand{\relenc}{\ensuremath{\rel_\enc}}
\newcommand{\releval}{\ensuremath{\rel_\eval}}

\newcommand{\seqi}[1]{\ensuremath{\left({#1}_i\right)_{i=1}^k}}
\newcommand{\seqii}[1]{\ensuremath{\left({#1}_i\right)_{i={k+1}}^m}}

\newcommand{\scheme}{\pcalgostyle{FHE}}
\newcommand{\fhe}{\ensuremath{\mathcal{E}}}
\newcommand{\zkp}{\ensuremath{\Pi}}
\newcommand{\schemedef}{\ensuremath{\mathcal{E} = (\kgen, \enc, \dec, \eval)}}
\newcommand{\phischeme}{\ensuremath{\vc\left[\scheme\right]}}
\newcommand{\proofscheme}{\ensuremath{\Pi}}
\newcommand{\proofschemedef}{\ensuremath{\Pi = (\setup, \prove, \verify)}}
\newcommand{\vc}{\pcalgostyle{VC}}
\newcommand{\vccpa}{\pcnotionstyle{VC\pcmathhyphen{}CPA}}
\newcommand{\vcvai}{\pcnotionstyle{VC\pcmathhyphen{}CVA1}}
\newcommand{\vcvaii}{\pcnotionstyle{VC\pcmathhyphen{}CVA2}}
\newcommand{\vcccai}{\pcnotionstyle{VC\pcmathhyphen{}CCA1}}
\newcommand{\vcccaiv}{\pcnotionstyle{VC\pcmathhyphen{}CCA1.5}}

\newcommand{\privccai}{\pcnotionstyle{PRIV\pcmathhyphen{}CCA1}}
\newcommand{\privcvaii}{\pcnotionstyle{PRIV\pcmathhyphen{}CVA2}}
\newcommand{\verifccai}{\pcnotionstyle{VER\pcmathhyphen{}CCA1}}
\newcommand{\verifcvaii}{\pcnotionstyle{VER\pcmathhyphen{}CVA2}}
\newcommand{\privatk}{\pcnotionstyle{PRIV\pcmathhyphen{}ATK}}

\let\privccaiTex\privccai
\renewcommand{\privccai}{\texorpdfstring{\privccaiTex}{PRIV-CCA1}}

\newcommand{\fspace}{\ensuremath{\mathcal{F}}}
\newcommand{\kspace}{\ensuremath{\mathcal{K}}}
\newcommand{\ptspace}{\ensuremath{\mathcal{M}}}
\newcommand{\ctspace}{\ensuremath{\mathcal{C}}}

\newcommand{\revek}{\ensuremath{\pcalgostyle{RevEk}}}

\newcommand{\indcpad}{\texorpdfstring{\pcnotionstyle{IND\pcmathhyphen{}CPA^D}}{IND-CPA-D}}

\newcommand{\findcpad}{\texorpdfstring{\pcnotionstyle{\fspace\pcmathhyphen{}IND\pcmathhyphen{}CPA^D}}{ℱ-IND-CPA-D}}
\newcommand{\ifindcpad}{\pcnotionstyle{1\pcmathhyphen{}\fspace\pcmathhyphen{}IND\pcmathhyphen{}CPA^D}}
\newcommand{\qfindcpad}{\pcnotionstyle{q\pcmathhyphen{}\fspace\pcmathhyphen{}IND\pcmathhyphen{}CPA^D}}
\newcommand{\findcpadi}{\pcnotionstyle{\fspace\pcmathhyphen{}IND\pcmathhyphen{}CPA1^D}}
\newcommand{\findcpadii}{\pcnotionstyle{\fspace\pcmathhyphen{}IND\pcmathhyphen{}CPA2^D}}
\newcommand{\ifindcpadi}{\pcnotionstyle{1\pcmathhyphen{}\fspace\pcmathhyphen{}IND\pcmathhyphen{}CPA1^D}}
\newcommand{\qfindcpadi}{\pcnotionstyle{q\pcmathhyphen{}\fspace\pcmathhyphen{}IND\pcmathhyphen{}CPA1^D}}
\newcommand{\ifindcpadii}{\pcnotionstyle{1\pcmathhyphen{}\fspace\pcmathhyphen{}IND\pcmathhyphen{}CPA2^D}}
\newcommand{\qfindcpadii}{\pcnotionstyle{q\pcmathhyphen{}\fspace\pcmathhyphen{}IND\pcmathhyphen{}CPA2^D}}

\let\indcpaTex\indcpa
\renewcommand{\indcpa}{\texorpdfstring{\indcpaTex}{IND-CPA}}

\let\indccaiTex\indccai
\renewcommand{\indccai}{\texorpdfstring{\indccaiTex}{IND-CCA1}}

\let\indccaiiTex\indccaii
\renewcommand{\indccaii}{\texorpdfstring{\indccaiiTex}{IND-CCA2}}

\newcommand{\indcvai}{\pcnotionstyle{IND\pcmathhyphen{}CVA1}}
\newcommand{\indcvaii}{\pcnotionstyle{IND\pcmathhyphen{}CVA2}}
\newcommand{\iindcvaii}{\pcnotionstyle{1\pcmathhyphen{}IND\pcmathhyphen{}CVA2}}
\newcommand{\qindcvaii}{\pcnotionstyle{q\pcmathhyphen{}IND\pcmathhyphen{}CVA2}}

\newcommand{\indccaiv}{\pcnotionstyle{IND\pcmathhyphen{}CCA1.5}}

\newcommand{\indrcca}{\pcnotionstyle{IND\pcmathhyphen{}RCCA}}
\newcommand{\indx}{\pcnotionstyle{IND\pcmathhyphen{}*}}

\newcommand{\iindccai}{\pcnotionstyle{1\pcmathhyphen{}IND\pcmathhyphen{}CCA1}}
\newcommand{\qindccai}{\pcnotionstyle{\ensuremath{q}\pcmathhyphen{}IND\pcmathhyphen{}CCA1}}

\newcommand{\indkhcca}{\pcnotionstyle{IND\pcmathhyphen{}KH\pcmathhyphen{}CCA}}
\newcommand{\iindkhcca}{\pcnotionstyle{1\pcmathhyphen{}IND\pcmathhyphen{}KH\pcmathhyphen{}CCA}}

\let\indkhccaTex\indkhcca
\renewcommand{\indkhcca}{\texorpdfstring{\indkhccaTex}{IND-KH-CCA}}

\newcommand{\indatk}{\pcnotionstyle{IND\pcmathhyphen{}ATK}}
\newcommand{\atk}{\pcnotionstyle{ATK}}

\newcommand{\ver}{\pcnotionstyle{VERIF}}
\renewcommand{\priv}{\pcnotionstyle{PRIV}}
\newcommand{\fpriv}{\pcnotionstyle{FPriv}}
\newcommand{\poe}{\pcnotionstyle{PoC}}

\newcommand{\cpa}{\pcnotionstyle{CPA}}
\newcommand{\cpad}{\pcnotionstyle{CPA^D}}
\newcommand{\ccai}{\pcnotionstyle{CCA1}}
\newcommand{\ccaii}{\pcnotionstyle{CCA2}}
\newcommand{\ccaiv}{\pcnotionstyle{CCA1.5}}
\newcommand{\cva}{\pcnotionstyle{CVA}}
\newcommand{\cvai}{\pcnotionstyle{CVA1}}
\newcommand{\cvaii}{\pcnotionstyle{CVA2}}

\newcommand{\zk}{\pcnotionstyle{ZK}}
\newcommand{\ks}{\pcnotionstyle{KS}}
\renewcommand{\ss}{\pcnotionstyle{SS}}

\newcommand{\cons}{\pcnotionstyle{COM}}

\newcommand{\ek}{\pckeystyle{ek}}
\newcommand{\rk}{\pckeystyle{rk}}
\newcommand{\anykey}{\pckeystyle{key}}

\newcommand{\expr}[3]{\ensuremath{\pcalgostyle{Expr}_{#1}^{#2}[#3](\secparam)}}  %
\renewcommand{\advantage}[2]{\ensuremath{\pcalgostyle{Adv}^{#1}[#2](\secpar)}} %
\newcommand{\orcl}{\ensuremath{\mathcal{O}}}

\newcommand{\id}{\pcalgostyle{id}}

\renewcommand{\set}[1]{\ensuremath{{\left\{#1\right\}}}}
\renewcommand{\sequence}[1]{\ensuremath{{\left[#1\right]}}}

\newcommand{\hlmath}[1]{\begingroup\setlength{\fboxsep}{0pt}\colorbox{gray!30}{\ensuremath{#1}}\endgroup}
\newcommand{\hltext}[1]{\begingroup\setlength{\fboxsep}{0pt}\colorbox{gray!30}{#1}\endgroup}
\newcommand{\clrmath}[1]{\ensuremath{\textcolor{ETHBlue}{{#1}}}}
\newcommand{\clrtext}[1]{\textcolor{ETHBlue}{#1}}

\renewcommand{\O}[1]{\ensuremath{\mathcal{O}\left(#1\right)}}

\newcounter{attack}[section]
\newenvironment{attack}[1][]{\refstepcounter{attack}\par\medskip
   \noindent \textbf{\sffamily Attack~\theattack. #1} \rmfamily}{\medskip}

\newcounter{construction}[section]
\newenvironment{construction}[1][]{\refstepcounter{construction}\par\medskip
   \noindent \textbf{ Construction~\theconstruction{}~(#1)} }{\medskip}

\newcommand{\circempty}{\Circle}
\newcommand{\circhalf}{\RIGHTcircle}
\newcommand{\circfull}{\CIRCLE}
\newcommand{\mymarker}{\circfull}

\def\tabularxcolumn#1{m{#1}}
\newcolumntype{C}[1]{>{\centering\arraybackslash}p{#1}}
\newcolumntype{Y}{>{\centering\arraybackslash}X}

\newcommand{\greyout}[1]{\textcolor{ETHGray!80}{#1}}

\setlist[description]{leftmargin=*}%
\renewcommand{\descriptionlabel}[1]{\hspace{\labelsep}\bfseries{#1}}

\newcommand{\qedhere}{\hfill\ensuremath{\square}}

\begin{acronym}
    \acro{ABE}{Attribute-Based Encryption}
    \acro{cca1}[\ccai]{non-adaptive Chosen Ciphertext Attack}
    \acro{cca2}[\ccaii]{adaptive Chosen Ciphertext Attack}
    \acro{cca1.5}[\ccaiv]{adaptive Chosen Decryption/Verification Attack}
    \acro{cpa}[\cpa]{Chosen Plaintext Attack}
    \acro{cva1}[\cvai]{non-adaptive Chosen Verification Attack}
    \acro{cva2}[\cvaii]{adaptive Chosen Verification Attack}
    \acro{FFT}{Fast Fourier Transform}
    \acro{FHE}{Fully Homomorphic Encryption}
    \acro{HA}{Homomorphic Authenticators}
    \acro{HE}{Homomorphic Encryption}
    \acro{indcca1}[\indccai]{Indistinguishability under non-adaptive Chosen Ciphertext Attack}
    \acro{indcca2}[\indccaii]{Indistinguishability under adaptive Chosen Ciphertext Attack}
    \acro{indcca1.5}[\indccaiv]{Indistinguishability under adaptive Chosen Ciphertext Decryption/Verification Attack}
    \acro{indcpa}[\indcpa]{Indistinguishability under Chosen Plaintext Attack}
    \acro{indcva1}[\indcvai]{Indistinguishability under non-adaptive Chosen Ciphertext Verification Attack}
    \acro{indcva2}[\indcvaii]{Indistinguishability under adaptive Chosen Ciphertext Verification Attack}
    \acro{indkhcca}[\indkhcca]{Keyed-Homomorphic Indistinguishability under adaptive Chosen Ciphertext Attack}
    \acro{LHE}{Leveled Homomorphic Encryption}
    \acro{MAC}{Message Authentication Code}
    \acro{ML}{Machine Learning}
    \acro{MPC}{Multi-Party Computation}
    \acro{NTT}{Number-Theoretic Transform}
    \acro{PIR}{Private Information Retrieval}
    \acro{PPT}{Probabilistic Polynomial-Time}
    \acro{PRF}{Pseudo-Random Function}
    \acrodefplural{PRF}{Pseudo-Random Functions}
    \acro{PSI}{Private Set Intersection}
    \acro{PSU}{Private Set Union}
    \acro{QRP}{Quadratic Ring Program}
    \acro{RLWE}{Ring-Learning With Errors}
    \acro{SNARG}{Succinct Non-interactive ARGument}
    \acro{SNARK}{Succinct Non-interactive ARgument of Knowledge}
    \acrodefplural{SNARK}{Succinct Non-interactive ARguments of Knowledge}
    \acro{SaaS}{Software as a Service}
    \acro{TEE}{Trusted Execution Environment}
    \acro{VANET}{Vehicular Ad-Hoc Network}
    \acro{VC}{Verifiable Computation}
    \acro{ZKP}{Zero-Knowledge Proof}
\end{acronym}
\section{Introduction}
\acf{FHE}, which enables computations on encrypted data, has recently emerged into practice.
Thanks to theoretical improvements, and optimizations in both software~\cite{Halevi2014-cb,Chen2017-xv,Al_Badawi2022-iv} and hardware implementations~\cite{Samardzic2021-yj,Geelen2022-tw,Feldmann2021-xs}, it is starting to see use in real-world deployments (e.g., the Microsoft Edge Password Monitor~\cite{Lauter2021-tk}).
Computing on encrypted data inherently requires malleable ciphertexts (e.g., the addition of two ciphertexts is also valid ciphertext).
However, this malleability also raises the issue of \emph{integrity}, as the server can deviate from the computation requested by the client.
This has obvious implications for correctness but can also have more severe consequences:
a malicious server can exploit the malleability of FHE to carry out key-recovery attacks~\cite{Fauzi2021-ip, Zhang2012-jk, Chillotti2016-zz, Chenal2015-bs, Chaturvedi2022-ff},
undermining the confidentiality of FHE.
So far, most work on FHE schemes and applications has chosen to side-step this issue by making assumptions on the setting and threat model.
However, as FHE is starting to be deployed to protect critical information, we must move beyond these assumptions to a threat model that can withstand real-world adversaries.
    
\fakeparagraph{Honest-but-Curious Assumption}
Historically, the FHE research community has extensively made use of the assumption that the server running an FHE application would be honest-but-curious, 
rather than actively malicious \cite{Fan2012-ip, Brakerski2014-qh, Chillotti2020-vx, Ducas2015-dx}. 
This assumption may be reasonable in some deployment scenarios (e.g., when FHE is used only to ensure regulatory compliance or when dealing with trusted institutions cooperating on their own data).
However, the necessity to trust the server to this extent is very limiting to the scope of application scenarios, since a violation of the assumption threatens not only correctness but also confidentiality.
In addition, even otherwise trusted parties can be compromised by malicious third parties, exposing this attack surface.
While FHE protects against passive attacks, a malicious or compromised server taking part in an FHE application can leverage this to undermine data confidentiality.
A class of attacks known as \emph{key-recovery attacks} exploits the interactive nature of real-world deployments to construct (partial) decryption oracles.
These exploit the fact that a server can craft a ciphertext that fails to decrypt correctly for certain secret keys, using the client's reaction or lack thereof as an oracle.
Practical key-recovery attacks have been developed for all major FHE schemes~\cite{Fauzi2021-ip, Zhang2012-jk, Chillotti2016-zz, Chenal2015-bs, Chaturvedi2022-ff}. 
Therefore, there is an urgent need to strengthen FHE to maintain strong guarantees in the context of these attacks.

\fakeparagraph{Existing FHE Integrity Approaches}
In order to remediate these attacks, a line of research has emerged that constructs more robust FHE schemes \cite{Boneh2007-sa, Loftus2012-ad, Li2016-ea, Lai2016-az, Emura2018-lv, Wang2018-qb, Emura2021-mr, Sato2022-lt} that achieve indistinguishability against chosen ciphertext attacks (\indccai).  
These schemes remain secure even in the presence of decryption oracles.
Unfortunately, many of these constructions assume the presence of cryptographic primitives even stronger than FHE and/or are too inefficient to implement in practice.
A different line of research focuses on achieving integrity for FHE; guaranteeing a function was correctly executed on the ciphertext while preserving the confidentiality of inputs~\cite{Gennaro2010-wo, Gennaro2013-zu, Catalano2013-qi, Fiore2014-zt, Fiore2020-op, Bois2021-qa, Ganesh2021-rq, Chatel2022-ei}.
While these are more concretely efficient than the constructions mentioned above, there is a significant gap between the assumptions made by existing work and the way state-of-the-art FHE schemes are used in practice. 
In particular, existing schemes can only tolerate  adversaries limited to verification oracles that are weaker than the decryption oracles present in most real-world settings. 

More broadly, the issue of maliciously-secure private computation has also been studied outside the domain of FHE.
For example, maliciously-secure secure \acf{MPC} has been studied extensively~\cite{Evans2018-ib}. 
However, most techniques do not transfer to the FHE setting since they rely on the interactive nature of MPC (e.g., cut-and-choose protocols).
In addition, there has been work studying how to construct verifiable computation with input privacy from other primitives~\cite{Goldwasser2013-zv}.
However, these approaches remain in the realm of theory, and recent work in this area builds almost exclusively upon FHE.

\fakeparagraph{Contributions}
This work is the first to consider integrity in the context of real-world FHE deployment settings, addressing the issue of FHE integrity holistically.
This paper aims to both 
highlight the dangers arising from the gap between existing notions and real-world scenarios,
and to propose efficient instantiations of a new robust notion for FHE  integrity that effectively addresses these challenges.
More concretely, this paper presents the following contributions:

\fakeparagraph{{\large \textcircled{\normalsize 1}} Analysis of FHE Integrity Constructions}
In this paper, we are the first to holistically analyze FHE integrity across the boundaries of different approaches.
Towards this, we unify the existing constructions into a set of general paradigms that form a taxonomy of FHE integrity approaches.
We then show how these notions fall short when considering how FHE is used in practice,
highlighting the mismatch between the setting assumed in the existing integrity literature and the settings used for the vast majority of FHE applications.
Finally, we show how this enables attacks on both correctness and confidentiality, even in the presence of existing integrity mechanisms.

\fakeparagraph{{\large \textcircled{\normalsize 2}} Formalizing Maliciously-Secure Verifiable FHE}
We define a new notion of integrity for FHE that captures real-world FHE deployment settings, addressing the issues we identified in our analysis.
Existing notions are usually a Frankensteinian combination of existing integrity notions
and ad-hoc confidentiality properties, interleaving FHE and integrity aspects. This makes them hard to reason about or extend.
In contrast, we present a natural clean-slate notion of verifiable FHE
that composes the standard notion of FHE with modular integrity properties. 
This allows our notion to adapt to the wide variety of FHE deployment settings we observe in practice.
We show how to generically construct our notion from a standard FHE scheme, commitments, and Zero-Knowledge Proofs (ZKPs).

\fakeparagraph{{\large \textcircled{\normalsize 3}} Instantiating Verifiable FHE in Practice}
We instantiate our notion of maliciously-secure verifiable FHE using a variety of different state-of-the-art ZKP systems.
In the process, we highlight a series of fundamental challenges in bringing together FHE and ZKP systems.
We investigate several approaches to bridge this gap and introduce a new optimization for emulating FHE ring arithmetic inside field-based ZKPs.
We evaluate our instantiations on a variety of different workloads 
and compare them to a hardware-attestation--based approach (FHE-in-TEE)  as point of comparison.
We show that verifiable FHE can be practical, but also highlight the need for future work on ZKP systems specifically designed for the unique characteristics of FHE.

\section{Background}
We briefly introduce Fully Homomorphic Encryption and important integrity techniques. 
We describe these concepts informally, highlighting properties and aspects relevant to our analysis, and refer to \Cref{app:furtherdef} for formal definitions of the concepts we rely on throughout the paper.
\vspace{-0.5em}

\subsection{Fully Homomorphic Encryption} \label{sec:background-fhe}
\vspace{-1em}
A \emph{homomorphic} encryption scheme allows meaningful computation to be performed on data while it is encrypted.
For example, in an additively homomorphic encryption scheme, $\dec(\enc(x+y)) = \dec(\enc(x)) \oplus \dec(\enc(y))$.
\emph{Fully} Homomorphic Encryption~\cite{Gentry2009-zu} offers both additive and multiplicative homomorphism.
For integer plaintext spaces, \emph{arithmetic circuits} with addition and multiplication gates allow the computation of any polynomial,
while for binary plaintext spaces, this emulates \emph{binary circuits} with \texttt{XOR} and \texttt{AND} gates and allows arbitrary computations to be performed on encrypted data.
Modern FHE schemes are based on the Learning with Errors (LWE)~\cite{Regev2009-am} or Ring Learning with Errors (RLWE)~\cite{Lyubashevsky2010-uo} hardness assumptions.
In this setting, carefully calibrated \emph{noise} is added to the encryption. 
This noise grows during computation, and once it crosses a certain threshold, decryption will no longer be correct.
FHE schemes address this by introducing \emph{ciphertext maintenance} operations that do not change the encrypted data but reduce the noise (growth) in a ciphertext.
FHE schemes are frequently used in \emph{leveled} mode, where they support a parameter-dependent fixed depth of computation before the noise overflows into the message.
Alternatively, \emph{bootstrapping}, which homomorphically refreshes the ciphertext, allows for arbitrarily deep computations.
However, bootstrapping is computationally expensive and, therefore, usually avoided in state-of-the-art FHE applications.

\vspace{-0.5em}
\subsection{Integrity Techniques}
\vspace{-0.5em}
We briefly introduce three key integrity techniques that have been used in the context of FHE.

\fakeparagraph{Message Authentication Codes}
A \ac{MAC} is a short unforgeable tag used to verify the authenticity and integrity of a message.
They are generated using a secret MAC key and then sent along with the corresponding message.
Note that verifying a MAC also requires access to the MAC key.
While MACs are usually intentionally non-malleable, \emph{homomorphic} MACs allow for some operations similar to homomorphic encryption. 

\fakeparagraph{Zero-Knowledge Proofs}
A \ac{ZKP} is a protocol that allows a \emph{prover} to convince a \emph{verifier} of the truth of a mathematical statement without revealing additional information.
Non-interactive ZKPs allow the prover to generate a \emph{proof} that the verifier can check independently, and there exist techniques to turn any ZKP non-interactive.
In the context of integrity, ZKPs can be used to show that $y=f(x,w)$ for a given $x,y$ and $f$ without revealing $w$.

\fakeparagraph{TEE Attestation}
\acp{TEE} are hardware components capable of isolating code running on them from the rest of the machine.
Code running on a TEE cannot be tampered with by other processes, even the operating system or hypervisor. 
TEEs are commercially available in commodity hardware provided by all major hardware vendors~\cite{sgx,Geater2015-do,amd-sev}.
TEEs are frequently used to provide confidentiality from either the server operators or other VMs running on the same hardware,  but they can also provide integrity through code \emph{attestation}, which generates a certificate that the outputs were generated through a correct execution of the program.

\section{Analysis of FHE Integrity Constructions}
\vspace{-0.5em}
\label{sec:survey}
In this section, we aim to answer the question: \emph{to what extent do the recent developments in FHE integrity address the needs of real-world FHE deployments, and where and why do they fall short?}
Recently, a number of approaches for FHE integrity have been proposed, covering a variety of settings and introducing a plethora of subtly different notions and properties.
In this section, we are the first to holistically analyze FHE integrity across the boundaries of the different approaches.
Towards this, we unify the existing constructions into a set of general paradigms that form a taxonomy of FHE integrity approaches.
This allows us to consider the strengths and weaknesses of the proposed notions independently of individual instantiations.
We then show how these notions fall short when considering how FHE is used in practice,
highlighting the mismatch between the setting assumed in the existing integrity literature and the settings used for the vast majority of FHE applications.
Finally, we show how this enables attacks on both correctness and confidentiality even in the presence of these integrity mechanisms.

\def\tabularxcolumn#1{m{#1}}
\begin{table*}[ptb]
\vspace{-1em}
\setlength{\tabcolsep}{0pt}
\begin{tabularx}{\linewidth}{l @{\hspace{0.5ex}}lYYYYY YY Y}
\toprule
\multicolumn{2}{c}{\multirow{2}{*}{}}
                        & \multirow{2}{*}{Ctxt. Maint.}  & \multirow{2}{*}{Circuit}   & \multirow{2}{*}{Server Inputs} & \multirow{2}{*}{Server Privacy} & \multirow{2}{*}{Approx. FHE}   & \multicolumn{2}{c}{Adversarial Model} & \multirow{2}{*}{Implementation}  \\
                        \cmidrule(lr){8-9}
  &  &  & & & & & Verif. Oracles & Dec. Oracles & \\
                        \midrule 
\multirow{3}{*}{MtE}%
&\cite{Gennaro2013-zu}  &    ?   & Any           & \circempty& \circempty    & \circempty& \circhalf & \circempty       & \circempty \\
&\cite{Catalano2013-qi} &   ?    & Any           & \circempty& \circempty    & \circempty& \circfull & \circempty     & \circempty \\
&\cite{Chatel2022-ei}  & \circfull & Any        & \circempty& \circempty    & \circempty& \circempty & \circempty         & \circfull \\
\midrule
EtM%
&\cite{Fiore2014-zt}    &  \circempty & Quadratic & \circempty& \circempty    & \circempty& \circempty & \circempty          & \circfull \\
\midrule
EaM%
&\cite{Li2018-ox}       &    ?   & Any           & \circempty& \circempty    & \circempty& \circempty & \circempty     & \circempty \\
\midrule
\multirow{3}{*}{ZKP}%
&\cite{Fiore2020-op} &\circempty & Any         & \circempty& \circfull     & \circempty& \circfull & \circempty     & \circempty \\
&\cite{Bois2021-qa}  &\circempty& LogspaceUnif & \circfull & \circfull     & \circempty& \circfull & \circempty     & \circempty \\
&\cite{Ganesh2021-rq}&\circhalf& Any         & \circfull & \circfull     & \circempty& \circempty & \circempty     & \circhalf \\
\midrule
TEE
&\cite{Natarajan2021-me} & \circfull& Any& \circfull & \circempty    & \circfull & \circfull & \circhalf     &\circhalf \\
\bottomrule
\end{tabularx}
\vspace{0.75\baselineskip}
\caption{Characteristics and limitations of existing FHE integrity paradigms and approaches. \\ A ? indicates insufficient details are given while $\circhalf$ indicates partial support. \ }
\label{tab:fhe-integrity}
\vspace{-2.5em}
\end{table*}

\subsection{Taxonomy of FHE Integrity Paradigms}
\vspace{-0.5em}
\label{sub:paradigms}%
We provide a complete taxonomy of all the literature on FHE integrity that we are aware of at the time of writing.
We analyze them according to the underlying techniques, grouping them into  Message-Authentication-Code--, Zero-Knowledge-Proof-- and  Attestation--based approaches.
We discuss the existing constructions at a level of abstraction that allows us to directly compare them and focus on their suitability for practical FHE deployments.
We provide a summary of our analysis in \Cref{tab:fhe-integrity}.

\fakeparagraph{Homomorphic Message Authentication Codes}
Message Authentication Codes (MACs) have long been used to ensure integrity for traditional symmetric-key encryption. 
In the context of FHE, special (fully) \emph{homomorphic} MACs are required, so that the server can combine valid MACs on the inputs to an FHE operation to a valid MAC of the output.
Existing constructions fall into three different paradigms based on how they combine the MAC and the underlying FHE scheme:
\emph{(i)}~Encrypt-and-MAC (EaM),
\emph{(ii)}~Encrypt-then-MAC (EtM), and
\emph{(iii)}~MAC-then-Encrypt (MtE).

In the Encrypt-and-MAC (EaM) paradigm, the initial MAC and FHE ciphertext are computed from the same plaintext and are then processed in parallel (but independently) to produce the output MAC and ciphertext pair.
Since the MAC in EaM is not encrypted under FHE, it must itself provide strong security, i.e., be semantically secure.
In addition, the MAC must offer the same homomorphic operations as the underlying FHE scheme.
Li et al. construct such an EaM scheme using multilinear maps, assuming a generic FHE scheme~\cite{Li2018-ox}.
While it describes how to support addition and multiplication operations,
it is not clear how this scheme can be extended to handle the complex ciphertext maintenance operations (e.g, relinearization) that are necessary for modern FHE schemes to achieve state-of-the-art efficiency.
In general, it is unclear how to expand the expressiveness of homomorphic MACs while maintaining the strong security guarantees required for the EaM approach.

The Encrypt-then-MAC (EtM) approach firsts encrypts the plaintext using FHE and then applies the MAC to the resulting ciphertext.
This removes the need for the MAC to preserve confidentiality, but 
the MAC does still need to be homomorphic with respect to operations on ciphertexts (including ciphertext maintenance operations).
Fiore et al. make use of this paradigm in~\cite{Fiore2014-zt}, instantiating their MAC using pairings. 
In order to enable this, they needed to introduce a homomorphic hash function to bridge the gap between FHE ciphertexts and the MACs, i.e., polynomial rings and pairing groups.
In order to achieve efficient verification, they also rely on amortized closed-form efficient PRFs~\cite{Benabbas2011-yr}.
However, the combination of these primitives limits the expressiveness of the resulting construction.
Specifically, it only supports quadratic circuits, i.e., circuits with at most one multiplication gate. 
In general, it is unclear whether it is possible to create MACs that support both arbitrarily deep circuits and complex operations on the ciphertexts.

Finally, the MAC-then-Encrypt (MtE) paradigm first computes a MAC over the plaintext and then encrypts the MAC-augmented plaintext under FHE.
This removes the need to support ciphertext maintenance operations,
as these do not affect the encrypted message.
Gennaro and Wichs~\cite{Gennaro2013-zu} provided one of the first FHE integrity construction based on this paradigm.
However, the construction is not efficiently verifiable, i.e., verifying the MAC requires recomputation that is as expensive as computing the original result. 
The work proposed potential ways to solve this issue, but did not instantiate a solution.
Catalano and Fiore~\cite{Catalano2013-qi} addressed the verification efficiency, but in turn their construction is limited to arithmetic circuits of a bounded depth.
More recently, Chatel et al.~\cite{Chatel2022-ei} have generalized these two approaches, providing the first FHE integrity scheme that can efficiently support arbitrary circuits and modern state-of-the-art schemes.

\fakeparagraph{Zero-Knowledge Proofs} 
\label{sub:CtP}
Zero-Knowledge Proofs are a natural primitive to explore for integrity constructions. 
In this approach, the server first computes the FHE circuit, storing intermediate ciphertext results, and then computes a \acf{SNARK} that asserts that the server knows an assignment of intermediate values to the circuit so that, for the given input, the circuit results in the output ciphertext.
In theory, any generic ZKP system could be used to generate this proof.
However, a trivial instantiation would introduce prohibitively large additional overhead in emulating the complex ring operations used in FHE schemes.
Recent work has therefore focused instead on developing ZKP systems tailored  to FHE.

One line of work uses (homomorphic) hashing to bring the size of FHE ciphertexts down into a range that can be handled more efficiently with ZKP techniques.
This includes the first \ac{SNARK} for FHE presented by Fiore et al.~\cite{Fiore2020-op} and follow-up work by Bois et al.~\cite{Bois2021-qa}.
However, the homomorphic hashing requirement limits this approach to simple schemes such as the BV scheme~\cite{Brakerski2011-xs} which does not feature the complex ciphertext maintenance operations that are necessary to achieve the practical efficiency enjoyed by state-of-the-art FHE schemes.
An alternative approach by Ganesh et al.~\cite{Ganesh2021-rq} instead focuses on constructing a generic ZKP system that natively operates on the rings used in FHE schemes.
This drastically improves the efficiency of proving the  ring operations that make up basic homomorphic operations such as addition and multiplication.
However, ciphertext-maintenance operations generally require either switching between different rings or non-ring operations such as rounding,
which are not supported by the current construction.
While such tailored approaches hold the promise of significant performance improvements, they cannot currently support efficient modern state-of-the-art FHE schemes.
Therefore, when we consider FHE integrity in practice in \Cref{sec:instantiation}, we will focus on generic ZKP systems and discuss how to efficiently instantiate it for FHE.

\fakeparagraph{Trusted Execution Environment Attestation}
\label{sec:tee-notions}
Trusted Execution Environments (TEEs) such as Intel SGX~\cite{sgx} can be used to provide confidentiality, but a series of attacks~\cite{Nilsson2020-mg,Fei2021-xb} has put their suitability for this task in question.
However, their integrity protections, i.e., their ability to \emph{attest} to the program running in the enclave, have so far mostly resisted practical attacks~\cite{Murdock2020-dl}.
Therefore, it is natural to augment the confidentiality properties of FHE with the integrity protections of TEEs by running FHE inside an enclave.
However, the computational complexity of FHE and especially the large sizes of ciphertexts and evaluation keys pose a challenge to TEEs, which are usually  more restricted in terms of memory and available computational power than the underlying untrusted hardware.
More fundamentally, they can only be employed in settings where the additional trust assumption on the specific hardware vendor is acceptable.
Natarajan et al.~\cite{Natarajan2021-me} present an FEE-in-TEE design and implementation, that, because of the ability of TEEs to express arbitrary computations, can easily support modern state-of-the-art FHE schemes with little to no required modifications.

\subsection{FHE Integrity in Practice} %
\label{sub:limitations}
In this section, we consider to what extent the assumptions and guarantees proposed in the existing FHE integrity literature fulfill the requirements of real-world FHE deployments.
We highlight the differences between the deployment settings assumed by the existing literature and the needs of real-world FHE deployments, uncovering significant mismatches.
In the following (Sections \ref{sub:correctness} and \ref{sub:confidentiality}), we discuss the implications of this mismatch for correctness and confidentiality.

\fakeparagraph{FHE Deployment Settings}
 The existing literature on FHE integrity assumes the \emph{outsourced computation} setting, where a client provides an encrypted input $x$ and a function (or circuit) $f$ to the server, which then computes $f(x)$ homomorphically and returns the encrypted result.
 While this setting is the most natural to define FHE in, it is not the only setting or even the most widespread one.
 In practice, the server has the ability to both choose the circuit to compute and to provide additional inputs. 
 This opens up a variety of important additional use cases that enable a form of two-party computation.
 For example, this can be used to offer privacy-preserving Machine Learning as a Service (MLaaS) where the server has a model that it wants to make available as a service and the client wants to receive a homomorphically computed inference on its private input~\cite{Wood2020-hz,Pulido-Gaytan2021-qu,Marcolla2022-ng}.
 More generally, we can categorize different types of FHE deployment by the aspects that must remain private.
 We assume that the client always has some secret inputs, but beyond that, we distinguish between settings with no server input, public (server) inputs, or private server inputs.
 We can also distinguish between settings with a public circuit $f$, and settings where the circuit is private to either the client or the server.
 In the following, we discuss the impact that these changes have on FHE integrity.
 
\fakeparagraph{(Private) Server Inputs}
    Public and private server inputs have significant implications for integrity, with the latter posing more fundamental challenges.
    Nevertheless, even public inputs prevent the use of some integrity approaches, such as MAC-based solutions~\cite{Gennaro2013-zu,Catalano2013-qi,Fiore2014-zt,Li2018-ox,Chatel2022-ei}.
    This is inherent in the concept of MACs, which only the client can generate for fresh messages.
    On the other hand, TEE- and ZKP-based approaches can usually support public inputs with straightforward extensions, even though these are not considered in the existing literature.
    Since these inputs are public, they can be transmitted to the client along with the result and proof (or attestation) of correctness, allowing them to complete the verification.
    Private server inputs, on the other hand, are more challenging because they raise fundamental questions about the meaning of correctness in their presence.
    For example, in the privacy-preserving MLaaS setting, it is not immediately clear what statements on the correctness of the computation can be made without revealing the model to the client.
    This highlights a fundamental gap between the \emph{application-level} correctness that we usually want to achieve and the \emph{circuit-level} correctness that existing integrity notions work on.
    We discuss this challenge in further detail in \Cref{sub:correctness}.
    More importantly, and perhaps unexpectedly, the ability to introduce private inputs opens up a new attack surface and allows a malicious server to undermine the \emph{confidentiality} of FHE.
    Intuitively, these issues arise because the private nature of the inputs allows the server to choose malformed inputs that `poison' the output in a way that will lead to a decryption failure at the client, which the server can observe through the client's reaction.
    While TEE- and ZKP-based approaches can technically be extended to support private inputs,  straightforward  extensions are ineffectual in preventing such attacks.  
    We discuss these issues in more detail in \Cref{sub:confidentiality}.

    \fakeparagraph{Private Circuits}
        A large number of FHE applications consider a setting where the circuits are public, which is appropriate for many applications either because the circuit definition is inherent in the task (e.g., Private Set Intersection) or because the majority of the private information is determined by inputs rather than the circuit (e.g., ML where architectures tend to be more standardized, but model parameters are the result of expensive data collection and training).
        The existing FHE integrity notions therefore almost always assume this public-circuit setting.
        However, more advanced scenarios could require the circuit to be private to either the client or the server. 
        In practice, FHE applications frequently assume that the circuit is provided by the server and not shared with the client.
        While it is technically possible to derive information ab, but the circuit from the resulting ciphertext by analyzing its noise, this is usually insufficient in practice to reveal significant additional information beyond that contained in the result itself.        
        Nevertheless, the FHE community has also developed techniques to achieve more formal circuit privacy guarantees, which use either bootstrapping~\cite{Kluczniak2022-gt,Ducas2016-ug,Bourse2022-jc} or \emph{noise flooding}~\cite{Gentry2009-gd} to essentially `standardize' the ciphertext before returning it to the client.
        Existing integrity notions, however, are incompatible with such an approach since they all require the circuit to be known to both parties.
        This can be addressed in theory by using a \emph{universal circuit} that takes the actual circuit as a (private) input.
        However, given the overhead of both the underlying FHE schemes and most integrity schemes, this will be infeasible in most scenarios.
        Note that this approach is also required when the circuit is private to the client, even when not using integrity systems.
        While currently impractical, future developments in FHE acceleration might make it feasible to use these techniques for simple, yet critical, high-value applications.

\subsection{Attacks on Correctness}
\label{sub:correctness}
    The ability of the server to provide inputs opens up a new attack surface
    since integrity issues can now arise not only from deviations in the computation but also from malicious inputs.
    In this setting, the server can produce `incorrect' results even when executing the circuit `correctly,' allowing them to produce a valid proof of correctness for these `incorrect' results.    
    We identify two different cases:
    first, the computation can result in an \emph{invalid ciphertext} that fails to decrypt correctly.
    Second, the computation can return a valid ciphertext but nevertheless fail to satisfy \emph{application-level} expectations.

\fakeparagraph{Invalid Ciphertexts}
    The noise inherent in FHE ciphertexts grows during computation and, for all but trivial circuits, must be managed carefully using ciphertext-maintenance operations.
    When the server inputs a ciphertext\footnote{This attack still works when the server provides a plaintext, since message size also impacts noise growth.} with larger noise than expected, the noise will overflow and garble the message. 
    As a result, decryption will no longer return the correct result.
    Note, however, that this is not necessarily observable by the client: the decryption will return an essentially random value, but without application-level constraints, this is indistinguishable from a correct result.
    Existing integrity notions do not consider this aspect since they are designed for the outsourced computation setting, where all inputs are from the client and can therefore be assumed to be well-formed.
    As a result, the server can provide an incorrect result together with a proof of `correctness,' clearly undermining the idea of integrity.
    More importantly, this also has severe implications on confidentiality, which we discuss in \Cref{sub:confidentiality}.

\fakeparagraph{Application-Level Correctness}
    Using existing integrity notions, a proof of `correctness' only guarantees that the result is the output of the specified circuit applied to the client's input and \emph{some} input provided by the server.
    When the server inputs are public, the client can verify that they are well-formed and meaningful.
    However, when the inputs are private, which is necessary for most applications in the two-party setting, this is no longer possible.
    This allows the server to affect the correctness of the result on an application level.
    For example,  consider a Private Information Retrieval (PIR) application, where a client wants to run a private query against a database stored on the server.
    The server could compromise application-level correctness by
    censoring individual entries in the database to further some malicious objective.
    For example, using existing integrity notions could hide specific Wikipedia articles or individual transactions from a blockchain history while still providing a proof of `correctness'.

\subsection{Attacks on Confidentiality (Key Recovery)}
\label{sub:confidentiality}
In addition to impacting correctness, the mismatch between existing notions and real-world requirements causes a considerably more serious issue:
a malicious server can exploit the interactive nature of practical FHE deployments to compromise confidentiality.
In fact, using \emph{key-recovery attacks}, a server could potentially recover the client's full secret key.
These attacks fall outside the scope of the semi-honest--server setting assumed by FHE schemes but should be addressed in a setting where the server is not trusted to execute the computation faithfully, such as that considered by integrity notions.
Existing integrity notions for FHE, however, fail to do so.

\fakeparagraph{Key Recovery Attacks}
Intuitively speaking, Key-Recovery Attacks exploit the fact that the decryption operation combines the ciphertext and the secret key.
While a decryption of a valid ciphertext will only ever output the encrypted message, a malformed ciphertext can result in (parts of) the secret key being returned instead.
For example, we briefly outline a simple key recovery attack by Chenal and Tang~\cite{Chenal2015-bs} against the BV scheme~\cite{Brakerski2011-xs}.
In BV, decryption is defined as $\dec_\sk(\ct) = \modb{\ct_0 + \ct_1 \cdot \sk}{t}$, where the inner operations are performed over $R_q$ (see \cite{Brakerski2011-xs} for details). 
When decrypting the special ciphertext $\ct = (0, 1)$, one trivially recovers the secret key $\dec_\sk(\ct) = \modb{0 + 1 \cdot \sk}{t} = \modb{\sk}{t} = \sk$ (under some assumptions on the parameters; we refer to \cite{Chenal2015-bs} for the details).
In this simple attack, the client can easily detect that this ciphertext has been maliciously crafted to be a (trivial) encryption of the secret key.
However, FHE also features encryptions of the secret key that are indistinguishable from standard ciphertexts.
For example, key-recovery attacks can take advantage of the \emph{evaluation keys} provided to the server in most schemes.
These are encryptions of (functions of) keys, which allow the server to perform crucial ciphertext maintenance operations (e.g., relinearization, key-switching, bootstrapping). 
If given access to a decryption oracle, an adversary could decrypt these evaluation keys and recover the secret key.
For example, we present how to exploit the relinearization key used in the  BGV scheme.
We recall that BGV \cite{Brakerski2014-qh} uses a relinearization key $\rk = (a\cdot \sk + t \cdot e + \sk^2, -a)$, which is a valid encryption of $\sk^2$ under $\sk$. 
Using a decryption oracle, we can recover $\sk^2$ and learn information about $\sk$. %

\fakeparagraph{Beyond Decryption Oracles}
 Full decryption oracles that would allow an adversary to exploit the straightforward attacks we discussed above are not commonplace in practice.
 However, more sophisticated attacks~\cite{Chillotti2016-zz} do not require such strong oracles and instead target decryption \emph{failure} oracles.
 These oracles arise when the server manipulates the result to have noise overflow, leading to an invalid decryption. 
 When the client detects and reacts to this unexpected result (e.g., by requesting a re-run of the computation or aborting further interactions),
 this leaks information to the adversary.
 Because most FHE application scenarios involve multiple queries to the server (e.g., MLaaS, PIR, etc), avoiding these oracles is frequently impossible in practice.
Common patterns in FHE applications frequently allow the server to have full control over the shape of the ciphertext result, even in the presence of existing integrity notions.
 For example, consider $f(x, w_1, w_2) = \langle x;~ w_1\rangle \,+\, w_2$, which appears when implementing a biased matrix-vector multiplication, a common operation in ML applications.
    The adversary can choose $w_1 = 0$ and an arbitrary $w_2$ as inputs, and prove that $f(x, w_1, w_2) = w_2$ for any $x$ provided by the client.
 However, these attacks usually do not even require full control of the output:
because state-of-the-art FHE applications are heavily optimized for efficiency, they usually have very limited additional noise capacity, so simply providing any outsized inputs is likely to cause a noise overflow.
 Since existing integrity notions only consider the correct execution of the operations in the circuit, they do not address this issue.
 Although some provide what appear to be strong guarantees of input privacy~\cite{Fiore2020-op,Bois2021-qa} that one might expect to exclude such attacks, even these do not prevent such attacks.
 This is because their privacy guarantees only hold against a much weaker adversary limited to verification oracles (i.e., informing the adversary whether an attempted proof passed verification).
 However, since decryption (failures) oracles are essentially impossible to avoid, any system trying to achieve FHE integrity in the real world must consider them. 
 In the following section, we, therefore, introduce a new robust notion of FHE integrity that is specifically designed to accommodate real-world deployment settings and threat models.

\section{Maliciously-Secure Verifiable FHE}
\label{sec:notions}
In this section, we define a new notion of integrity for FHE.
Specifically, it captures real-world FHE deployment settings, 
including those with actively malicious server inputs.
Existing notions are usually a Frankensteinian combination of an existing integrity notion (e.g., MAC unforgeability~\cite{Li2018-ox} or Verifiable Computation~\cite{Fiore2014-zt}) and ad-hoc confidentiality properties.
As a result, they generally interleave FHE and integrity aspects, which makes them hard to reason about or to extend.
In contrast, we present a natural notion of verifiable FHE
that cleanly composes the standard notion of FHE with integrity properties.
Whereas existing notions are monolithic, our modular framework allows us to easily define extensions of the core notion.
This enables us to support a wide range of FHE deployment settings with varying requirements.
We first present the core definition of a maliciously-secure verifiable FHE scheme (vFHE) and present the soundness, completeness, correctness, and security properties it must fulfill.
Then, we describe how to extend this notion with privacy for the server inputs, as well as with  input predicates.
Finally, we discuss how to generically construct a vFHE scheme from an \indcpa{}  secure FHE scheme, commitments, and a compatible ZKP system before studying how to instantiate this construction efficiently in the next section (\Cref{sec:instantiation}).

\subsection{Defining Maliciously-Secure Verifiable FHE}

One of the key insights of our notion is that we consider a stronger and arguably more realistic threat model.
Existing notions result in an inconsistent view of the server, 
which is both expected to deviate from the computation (otherwise, integrity notions would not be necessary)
yet, at the same time, assumes that the server will not use these deviations to exploit common decryption (failure) oracles that arise in real-world FHE to undermine confidentiality.
This can result in a false sense of security, leading users of constructions achieving these notions to believe that they have stronger protections than these notions in fact offer.
Because of this inconsistency, we believe that the natural threat model beyond the semi-honest server assumption should be an actively malicious server with full access to a decryption oracle.
In addition, we also consider an unbounded verification oracle in order to give similar strength to the integrity guarantees.

While there is a natural definition for confidentiality in this threat model (i.e., \indccai{} security\footnote{FHE cannot achieve \indccaii{} by its nature.}),
extending existing FHE integrity notions with this guarantee is non-trivial 
because these notions interleave aspects of integrity and confidentiality.
As a result, formalizing the interactions that arise between these (and their oracles) would be challenging and prone to errors.
We instead detangle confidentiality and integrity guarantees.
For example, we split correctness $\left(\dec(\enc(m))=m\right)$ and completeness (i.e., verifiability) which are usually combined in existing notions.
This allows us to reason about constructions more cleanly, and also allows us to easily define extensions of our core notion.
For example, we are the first to extend integrity notions to \emph{approximate} FHE~\cite{Cheon2017-ru}, where correctness holds only approximately $\left(\dec(\enc(m)) \approx m\right)$.
Because we split correctness and completeness, changing between verifiable FHE and verifiable approximate FHE has no knock-on-effects on the remainder of the notion.
For space reasons, we present the exact notion of correctness here and refer to \Cref{app:epscorrectness} for a discussion of the approximate notion.

\begin{definition}[Maliciously-Secure Verifiable FHE]
\label{def:vfhe}
A \emph{maliciously-secure verifiable FHE}  (vFHE) scheme is a tuple $(\kgen, \enc, \eval, \verify, \dec)$ of PPT algorithms:
\begin{itemize}
\item $\kgen(\secparam, f) \to (\pk, \sk)$ 
\item $\enc_\anykey(x) \to (c_x, \tau_x)$ where \anykey{} is either\footnote{When \anykey{} is \pk{}, the scheme achieves \emph{public delegatability}, which is natural for FHE since all modern schemes offer public-key encryption.} \pk{} or \sk{}
\item $\eval_{\pk}(c_x) \to (c_y, \tau_y)$ where  $y = f(x)$
\item $\verify_\sk(c_y, \tau_x, \tau_y) \to b$, the client accepts if $b = 1$ and rejects otherwise
\item $\dec_{\sk}(c_y) \to y$
\end{itemize}

\noindent
The scheme must satisfy the 
\emph{correctness}, %
\emph{completeness}, %
\emph{soundness}, %
and \emph{security} %
properties defined below in the context of our adversary model.
\end{definition}

Note that this notion already supports \emph{public} inputs on the server, 
modeling them as part of the function $f$.
We discuss how to extend it to support private server inputs in the next section.
Note also that $f$ is public by default, and we discuss how to extend this notion to support circuit privacy in a following section.
We present our notion with \emph{designated verifiability}, i.e., some secret key material might be required to verify a ciphertext.
We believe that this is natural for the FHE setting, where only the secret key holder can decrypt a ciphertext and therefore has any benefit from verifying it.
However, our notion can  be trivially adapted to \emph{public verifiability} if desired.
Note that even in a threshold FHE or multi-key FHE setting, designated verifiability is sufficient, since the key holders can extend the MPC protocol they run to decrypt to also realize verification.

\fakeparagraph{Adversary Model}
    We assume a PPT adversary with access to an unbounded verification and decryption oracle
    $\orcl_\dec(c_x, \tau_y, \tau_z):=b$ which returns $b =  \bot$ if $\verify_\anykey(c_x, \tau_y, \tau_z) = 0$, and $b = \dec_\sk{}(c_x)$ otherwise\footnote{We note that this decryption oracle returns $\bot$ for fresh ciphertexts. However, the adversary can still use the oracle for a fresh ciphertext by generating a output tag for the ciphertext output by the identity function.}. 
    In addition, the adversary has access to the usual encryption oracle $\orcl_\enc$, which we omit for brevity.

\begin{definition}[Correctness]
\label{def:correctness}
A scheme is correct if any honest computation will decrypt to the expected result.
More formally, a scheme is \emph{correct} if for all functions $f$, and for all $x$ in the domain of $f$: 
\begin{equation*}
\condprob{
\begin{gathered}
\dec_\sk(c_y) = f(x)
\end{gathered}
}{
\begin{gathered}\
(\pk, \sk) \gets \kgen(\secparam, f)\\
(c_x, \tau_x) \gets \enc_\anykey(x) \\
(c_y, \tau_y) \gets \eval_\anykey(c_x)
\end{gathered}
} = 1
\end{equation*} 

\end{definition}

\begin{definition}[Completeness]
\label{def:completeness}
A scheme is complete if \verify{} will always accept an honestly computed result.
More formally, a scheme is \emph{complete} if for all functions $f$, and for all $x$ in the domain of $f$: 
\begin{equation*}
\condprob{
\begin{gathered}
\verify_\sk(c_y, \tau_x, \tau_y) = 1 
\end{gathered}
}{
\begin{gathered}
(\pk, \sk) \gets \kgen(\secparam, f)\\
(c_x, \tau_x) \gets \enc_\anykey(x) \\
(c_y, \tau_y) \gets \eval_\anykey(c_x) \\
\end{gathered}
} = 1
\end{equation*} 
\end{definition}

\begin{definition}[Soundness]
\label{def:soundness}
A scheme is sound if the adversary cannot make \verify{} accept an incorrect answer. 
Formally, a  scheme is \emph{sound} if for any PPT adversary \adv{} and any function $f$ the following probability is negligible in the security parameter \secpar: 

\begin{equation*}
\condprob{
\begin{gathered}
\verify_\sk(c_y, \tau_x, \tau_y) = 1 \\
\land \\
\dec_\sk(c_y) \neq f(x)
\end{gathered}
}{
\begin{gathered}
(\pk, \sk) \gets \kgen(\secparam, f)\\
x \gets \adv^{\orcl_\enc, \orcl_\dec}(\pk) \\
(c_x, \tau_x) \gets \enc_\anykey(x) \\
(c, \tau_y) \gets \adv^{\orcl_\enc, \orcl_\dec}(c_x, \tau_x) \\
\end{gathered}
}
\end{equation*} 
\end{definition}

\begin{definition}[Security]
\label{def:security}
We extend the definition of \indccai{} to our notion. 
Note that we do not require an evaluation oracle since $\eval_\pk$ is public and can be executed by the adversary directly.
Formally, a scheme is \emph{secure} if for any PPT adversary \adv{} and any function $f$ the advantage $\advantage{\indccai}{\adv} = 2 \left| \prob{b = \widehat{b}} - \frac{1}{2}\right|$ of the attacker in the following game is negligible in the security parameter $\secpar$:
\begin{center}
\pcsetargs{headlinesep=-0.5ex}
\procedure{\indccai{} for vFHE}{
(\pk, \sk) \gets \kgen(\secparam, f) \\
(\pt_0, \pt_1) \gets \adv^{\orcl_\enc, \orcl_\dec}(\secparam, f, \pk) \\
(c^\ast, \tau^\ast) \gets \enc_\pk(\pt_b) \\
\widehat{b} \gets \adv^{\orcl_\enc}(c^\ast)
}
\end{center}
\end{definition}

\subsection{Server Privacy} %

Verifiable FHE as defined in \Cref{def:vfhe} addresses key-recovery attacks and ensures that client inputs are protected as expected.
However, it does not address the privacy of server inputs and we therefore provide an extension of the notion which can be used in settings where formal guarantees for server privacy are required.
In the existing FHE literature, the notion of \emph{circuit privacy} addresses both the privacy of the function and any server inputs.
It requires that the result of a computation is indistinguishable from a fresh encryption (with the same parameters).
However, this is a stronger guarantee than what we require here since we address circuit privacy as a separate property in the next section.
In addition, the traditional definition might be hard to achieve in practice, given that the integrity tags $\tau$ of a fresh encryption and a computation output might be substantially different, making distinguishing them trivial.

We instead define our notion as vFHE plus a new property we term ``Server Privacy''.
Informally speaking, it requires that the result of an evaluation reveals nothing 
to  the client beyond the output of the function.
Existing work that considers hiding server inputs does not explicitly state the threat model for this setting.
We address this issue and define an adversary model that considers the client as the adversary.
For example, considering the perspective of the client, the indistinguishability has to hold even when the adversary has access to the secret key.
In addition, we must assume that the client generates keys and encrypts honestly, i.e., we assume a semi-honest client. 
We could strengthen the threat model at the cost of requiring proofs of correct key generation and encryption.
However, we believe that this would be prohibitively expensive and not appropriate for most settings.
We note the parallels between this setting and generic 2-party MPC.
However, in MPC, parties usually have equal protection.
Here, the guarantee offered to the client is stronger, since the server never learns the output of the function and therefore has \emph{no} information on the client input, not even that which would be derivable from the function output.

\begin{definition}[vFHE with Server Privacy]
\label{def:vfhesp}
A \emph{malicously-secure verifiable FHE scheme with server privacy} is a tuple $(\kgen, \enc, \verify, \dec, \eval)$ of PPT algorithms:
\begin{itemize}
\item \kgen, \enc, \verify, \dec{} are defined as in \Cref{def:vfhe}, 
\item $\eval_{\pk}(c_x, w) \to (c_y, \tau_y)$ where  $y = f(x,w)$ \\ for private server inputs $w$
\end{itemize}

\noindent
The scheme must satisfy (slightly modified versions of) the
\emph{correctness}, %
\emph{completeness}, %
\emph{soundness},
and \emph{security} %
properties defined above:
\begin{itemize}
    \item For correctness (\Cref{def:correctness}) and completeness (\Cref{def:completeness}) only trivial syntactic changes to accommodate the modified \eval{} are required.
    \item For soundness (\Cref{def:soundness}), $\dec_\sk(c_y) \neq f(x)$ is replaced by $\nexists{} w$ s.t. $\dec_\sk(c_y) = f(x,w)$.
    Informally, this means that if \verify{} accepts, then there exists at least one set of private server inputs that satisfies the circuit.
    \item The definition of security (\Cref{def:security}) remains unchanged, but the scheme must additionally satisfy the \emph{server privacy} property defined below.
\end{itemize}
\end{definition}

\begin{definition}[Server Privacy]
    Informally, a scheme offers server privacy when \eval{} reveals nothing about the server inputs beyond what can be learned from the output.
    We assume a semi-honest PPT adversary with access to all keys, including \sk.
    Formally, a scheme offers \emph{server privacy} when the following are (statistically) indistinguishable for all $w,w'$:
    \[ (f(x,w), \eval_\pk(c_x,w)) \approx (f(x,w'), \eval_\pk(c_x,w')) \]
\end{definition}

\subsection{Circuit Privacy}
When considering circuit privacy, there are two natural versions of this property:
We can have either the server or the client provide the (private) circuit.
While applications relying on this are currently rare in practice, certain critical applications could require formal guarantees of this nature.
While the FHE literature has defined the notion of \emph{circuit privacy} from the very beginning, existing integrity notions generally did not consider this aspect.
This seems initially justified in the case of a server-provided circuit since the resulting correctness guarantees are essentially meaningless.
However, this is only true when ignoring decryption oracles (as existing notions do).
In the presence of decryption oracles, it remains meaningful to ensure the well-formedness of ciphertexts to prevent key-recovery attacks.
Client-provided circuit privacy, on the other hand, does not have a correspondence in the traditional FHE literature, nor has it appeared in existing integrity notions.
However, of the two, it is arguably the more interesting in the context of integrity since one can extend the privacy guarantees to the circuit while  maintaining the same strong correctness guarantees.
Because our notion separates the concepts of server input privacy and circuit privacy, we can easily define both client and server versions of the latter.
Both are very similar and, in the following, we define circuit privacy only from the perspective of the client and refer to \Cref{app:furtherdef} for the server circuit privacy definitions.
We define circuit privacy as an extension of the core vFHE notion (\Cref{def:vfhe}), however, it can be trivially defined as an extension of vFHE with server privacy, too.

\begin{definition}[vFHE with Client Circuit Privacy]
A \emph{malicously-secure verifiable FHE scheme with client circuit privacy} is a tuple $(\kgen, \enc, \verify, \dec, \eval)$ of PPT algorithms:
\begin{itemize}
\item $\kgen(\secparam, \mathcal{F}) \to (\pk, \sk)$ 
\item \enc, \dec, and \verify{} are defined as in \Cref{def:vfhe}
\item $\eval_{\pk}(c_x, c_f) \to (c_y, \tau_y)$ where  $y = f(x)$ for $f \in \mathcal{F}$
\end{itemize}

\noindent
The scheme must satisfy the 
\emph{correctness}, %
\emph{completeness}, %
\emph{soundness}, %
and \emph{security} %
properties defined above (with trivial syntactic changes as necessary).
In addition, it must satisfy the \emph{client circuit privacy} property defined below.
\end{definition}

\begin{definition}[Client Circuit Privacy]
    Informally, a scheme offers client circuit privacy when \eval{} reveals nothing about the function.
    Formally, a scheme offers \emph{client circuit privacy} when, for all $f,f' \in \mathcal{F}$, the following are (statistically) indistinguishable:  
    \[ \eval_\pk(c_x,c_f) \approx \eval_\pk(c_x,c_{f'}) \]    
\end{definition}

\subsection{Input Predicates}
Guarantees about the correct execution of a circuit are usually not sufficient to guarantee desirable application-level properties.
Arguably, there is little difference to the client between a `correct' execution of the circuit on `incorrect' inputs and an incorrect execution of the circuit.
While some amount of possible deviation is inherent in any 2-party computation,
a rich history of MPC and ZKP applications has shown how to use input checks to ensure that a computation achieves higher-level properties~\cite{Evans2018-ib}.
For example, in Machine Learning as a Service (MLaaS), a client can run several validation queries on inputs with known labels to ensure the model has sufficient accuracy before querying it with real inputs.
However, in the privacy-preserving MLaaS setting, where the model weights are private to the server, the server could switch out or degrade the model without the client noticing.
This can be solved efficiently by requiring the server to commit to the model weights before the validation and then ensuring that the provided model weights match the commitment as part of each query.
Alternatively, the client might want to ensure that the server inputs lie inside a valid range that is a subset of the possible plaintext space,
and, more generally, we can consider arbitrary predicates on the server inputs that must hold for the client to accept the result.

We note that one could, in theory, integrate such input checks into the function $f$.
However, this would require evaluating them under FHE and potentially cause issues when trying to combine this with the notion of circuit privacy.
We define our notion as an extension of vFHE with private server inputs, 
as it only really applies in this setting:
public inputs  are modeled as part of the circuit and can be trivially checked for consistency by the client. 
We model the input checks as a predicate $\phi: w \mapsto \bin{}$ and extend the notion as follows:

\begin{definition}[vFHE with Input Predicates]
\label{def:vfheip}
A \emph{malicously-secure verifiable FHE scheme with client circuit privacy} is a tuple $(\kgen, \enc, \verify, \dec, \eval)$ of PPT algorithms as defined in \Cref{def:vfhesp}.

\noindent
The scheme must satisfy (slightly modified versions of) the
\emph{correctness}, %
\emph{completeness}, %
\emph{soundness},
and \emph{security} %
properties defined above:
\begin{itemize}
    \item For correctness (\Cref{def:correctness}) and completeness (\Cref{def:completeness}), the probability is also conditional on $\phi(x) = 1$.
    \item For soundness (\Cref{def:soundness}), the conditionals remain the same, but we consider the probability of 
    \[\verify_\sk(c_y, \tau_x, \tau_y) = 1 \ \land \]
    \[\left( \dec_\sk(c_y) \neq f(x)  \lor \phi(w) = 0 \right)\]
    Informally, this means that if \verify{} accepts, the predicates must have also held.
    \item The definition of security (\Cref{def:security}) remains unchanged.
\end{itemize}
\end{definition}

\subsection{Generic Construction}
\label{sec:generic-constructions}

We show how to generically construct a maliciously secure verifiable FHE scheme (\Cref{def:vfhe}) from a standard FHE scheme and a generic ZKP system.
In order to achieve security in this malicious setting (especially when considering private server inputs) we need to combine a proof of circuit correctness (i.e., of correct computation) with well-formedness checks on the inputs that ensure the result will be safe to decrypt.
This requires us to make the concept of \emph{valid} decryptions explicit, 
which will usually have an application-dependent component.
For example, a logit vector returned by a machine learning model should consist of probabilities that sum to one.
Therefore, we do  not hard-code a concept of validity into our construction, but instead, define it relative to a set of functions that return valid results (for a certain range of inputs).

We build our construction from a standard \indcpa{} FHE scheme, which includes all modern state-of-the-art schemes.
While it might initially appear more attractive to utilize an \indccai{} FHE scheme, since our security notion is similarly defined,
this is not ideal.
Comparatively few \indccai{} FHE schemes are known~\cite{Canetti2017-by, Wang2018-qb, Li2016-ea} and the existing schemes are computationally much more expensive than \indcpa{} schemes.
This is because existing \indccai{} schemes either already explicitly include integrity protections~\cite{Canetti2017-by} or rely on powerful primitives that, in turn, would require integrity-like protections~\cite{Wang2018-qb,Li2016-ea} to realize.
As a result, starting our construction with an \indccai{} scheme and then adding integrity protections for circuit correctness would lead to duplicated efforts when instantiating the construction in practice.

Note that constructing a scheme for the core vFHE definition is comparatively straightforward, as this setting only features public server inputs:
for any $f$ (which includes the public inputs) that results in a valid ciphertext for any freshly encrypted client inputs, we can combine an \indcpa{} secure FHE scheme with a MAC-then-Encrypt scheme such as Chatel et al.~\cite{Chatel2022-ei} or a generic ZKP proof of circuit correctness.
Therefore, we instead focus on showing how to achieve the more complex notion of verifiable FHE with private server inputs, where we must actively counteract the possibility of key-recovery attacks.
We discuss this construction below and refer to \Cref{app:generic-constructions} for constructions of the core vFHE definition and the other extensions, including approximate correctness, circuit privacy, and input predicates.
Finally, we discuss concrete instantiations of these constructions in the following section (\Cref{sec:instantiation}).

\begin{construction}[vFHE with Private Server Input]
\begin{itemize}
    \item Let \fhe{} = (\fhe.\kgen, \fhe.\enc, \fhe.\dec, \fhe.\eval) be an \\ \indcpa{} secure FHE scheme with (FHE) circuit privacy (See \Cref{app:furtherdef} for formal definitions).
    \item Let $\mathcal{F_{\mathcal{X},\mathcal{W}}}$ be a set of functions, with  $f \in \mathcal{F}$ if
    for all $x \in \mathcal{X}$ and $w \in \mathcal{W}$, $\fhe.\dec_{\fhe.\sk}(\fhe.\eval(c_x, w))$ results in a valid plaintext\footnote{Note that the converse need not hold; a practical realization could rely on one-sided bounds on noise growth.} (which can be a subset of all possible plaintext). 
    \item Let \zkp{} = (\zkp.\kgen, \zkp.\prove, \zkp.\verify) be a generic zero-knowledge proof (e.g., a zkSNARK) for $\text{ZKPPoK}\left\{ w : c_y = \fhe.\eval_{\pk_\fhe}(f, c_x,w)
    \land \ w \in \mathcal{W} \right\}$.
\end{itemize}

\noindent
For  $f \in \mathcal{F}$ , we construct a vFHE scheme with private server inputs $(\kgen, \enc, \eval, \verify, \dec)$ satisfying \Cref{def:vfhesp} from these building blocks as follows:
\begin{itemize}
\item $\kgen(\secparam, f) \to \left((\pk_\fhe, \pk_\zkp), (\sk_\fhe, \sk_\zkp)\right)$, where \\
$(\pk_\fhe, \sk_\fhe) = \fhe.\kgen(\secparam, f)$ and\\
$(\pk_\zkp, \sk_\zkp) = \zkp.\kgen(\secparam, f)$
\item $\enc_\pk(x) \to \left(c_x, \tau_x\right)$, where $\tau_x = c_x$ and $c_x = \left(\fhe.\enc_{\pk_\fhe}(\pt)\right)_{\pt \in x}$ for $x \in \mathcal{X}$
\item $\eval_\pk(c_x, w) \to (c_y, \tau_y)$, where $c_y = \fhe.\eval(f,c_x,w)$ and $\tau_y = \zkp.\prove(c_y, \tau_x,w)$
\item $\verify_\sk(c_y, \tau_x, \tau_y) = \zkp.{\verify_\zkp.\sk}\left((\tau_y, \tau_x)\right)$. 
\item $\dec_\sk(c_y) \to y$, where $y = \fhe.\dec_{\sk_\fhe}(c_y)$
\end{itemize}
\end{construction}

\begin{proof}
The correctness of our construction reduces  to the correctness of the FHE scheme, 
and the completeness reduces to the completeness of the ZKP scheme. 
Security can be reduced to the semantic security of the FHE scheme (because we assume $f \in \mathcal{F}_{\mathcal{X},\mathcal{W}}$ and enforce $w \in \mathcal{W}$ 
 and correct execution through the ZKP, and $x \in \mathcal{X}$ in the encryption).
Soundness follows directly from the soundness of the \ac{ZKP},
while server privacy is achieved via the circuit privacy (as defined in FHE) of the FHE scheme and through the zero-knowledge property of the ZKP.
We refer to the extended version of this paper~\cite{extended-version} for a more formal proof.
\end{proof}
\section{Instantiating Verifiable FHE in Practice} 
\label{sec:primitives}
\label{sec:instantiation}
In this section, we instantiate our notion of maliciously-secure verifiable FHE using state-of-the-art FHE and ZKP schemes.
We highlight a series of challenges in bringing together modern FHE and ZKP systems, including the mismatch between the large polynomial rings used in most state-of-the-art FHE schemes and the integer fields used in the vast majority of ZKP systems.
We investigate several approaches to bridge this gap and introduce a new optimization for emulating ring arithmetic inside ZKPs.

We consider a wide range of ZKP systems and identify four promising candidates that are best suited to the characteristics of FHE verification.
In addition, we also discuss how to instantiate our notion with hardware attestation primitives, introducing an optimization that allows us to accelerate FHE-in-TEE by a factor of two over the existing state-of-the-art for multiplications.
We evaluate our ZKP- and TEE-based instantiations for a variety of different workloads going far beyond the type of circuits that existing FHE integrity notions can express.

\subsection{Verifiable FHE via ZKP}
In the following, we discuss how to instantiate our construction for verifiable FHE with private server inputs (c.f. \Cref{sec:generic-constructions}).

\fakeparagraph{Bridging FHE and ZKP}
The areas of FHE and ZKP have been maturing mostly independently,
and  state-of-the-art ZKP systems have primarily been tailored to applications that share few characteristics with FHE.
In addition, FHE computations are inherently large and complex, making proofs non-trivial.
Most of this complexity arises from the advanced ciphertext maintenance operations used by state-of-the-art schemes.
As a result, previous work on FHE integrity frequently chose to use simple schemes such as the BV scheme~\cite{Brakerski2011-xs} to avoid this complexity.
However, the applicability of these schemes is limited in practice because of their prohibitive overhead.
We instead choose to target modern state-of-the-art FHE schemes which offer the performance necessary to  realize real-world FHE applications.

FHE schemes fall into two main families, with the LWE-based FHEW~\cite{Ducas2015-dx} and TFHE~\cite{Chillotti2020-vx} schemes focusing primarily on evaluating binary circuits, while RLWE-based schemes such as BGV~\cite{Brakerski2014-qh}, B/FV~\cite{Brakerski2012-ox}\cite{Fan2012-ip} and CKKS~\cite{Cheon2017-ru}) focus on arithmetic circuits.
Initially, LWE-based schemes might appear promising since they use smaller ciphertexts and offer faster computation.
However, efficient implementations of schemes in this family usually make heavy use of floating-point operations, which would introduce significant overhead in the ZKP proofs.
The RLWE schemes, in turn, feature larger ciphertexts but also offer a powerful form of data parallelism~\cite{Smart2014-uh} that is at the core of most state-of-the-art FHE results.
While our construction can be instantiated with any scheme, we select BGV because it is amenable to integer-based ZKP and requires the least amount of non-arithmetic operations to realize its ciphertext-maintenance operations.

The RLWE setting introduces a fundamental mismatch between the \emph{ring} based FHE computation and the mostly \emph{field} based ZKP systems.
Specifically, BGV uses rings of the form $R_q := \mathbb{Z}_q\left[X\right]/\langle X^N+1\rangle$, i.e., polynomials with degree up to $N$ (usually $N>2^{13}$) and coefficients in $\mathbb{Z}_q$.
Most ZKP systems, on the other hand, mostly use large prime fields (i.e., $\mathbb{Z}_p$ where $p$ is usually a 254-bit prime).
Note that multiplying polynomials efficiently requires converting them into a form that allows element-wise multiplication via the Number Theoretic Transform (NTT), a discrete analog to the Fast Fourier Transformation (FFT).
As a side effect, this also removes the need to explicitly compute the reduction modulo $X^N+1$.
However, we must still take care to achieve modular reduction with regard to the coefficient modulus $q$.

\fakeparagraph{Matching the Coefficient Modulus}
Existing work mostly assumes that it is possible to instantiate the FHE  and ZKP schemes so that the ZPK field modulus $p$ is the same as the FHE coefficient modulus $q$.
While FHE already needs to support a wide range of ciphertext moduli since this is an application-dependent parameter, most ZKP systems offer less flexibility.
There is also an inherent tension between the FHE and ZKP parameters here, since FHE security reduces with larger $q$ (requiring increases in $N$ to compensate) while many ZKP systems rely on elliptic curves becoming more secure as $p$ increases.
As a result, while it is possible to instantiate FHE with a matching coefficient, this results in an unnecessarily inefficient FHE scheme, especially for smaller circuits.
On the other hand, for larger FHE circuits, $q$ might outgrow the ZKP modulus.
More fundamentally, this approach does not support the \emph{modulus switching} used by BGV and other state-of-the-art FHE schemes to manage noise.
Both of these issues can be addressed by using the Chinese Remainder Theorem (CRT) to decompose the coefficient modulus $q$ into $L$ moduli $q_1, \ldots, q_L$, working on each \emph{Residue Number System} (RNS) limb independently.
This is already frequently used in FHE implementations to improve performance on existing hardware, as computing ten 60-bit operations is significantly cheaper than computing one 600-bit operation.
However, even with the RNS approach, matching ZKP and FHE parameters remains difficult and inherently limited to trivial circuits, as it cannot express more complex ciphertext maintenance operations.

\fakeparagraph{Emulating the Coefficient Modulus}
Removing the need to match FHE and ZKP parameters allows more efficient instantiations of both, but requires us to emulate the FHE coefficient modulus inside the ZKP field.
This can be done by explicitly computing the modulus (mod $q$) after each arithmetic operation, which is comparatively cheap.
However, we also need to prove the correctness of the result, which requires two expensive range proofs.
We introduce an optimization that reduces the need to perform these expensive modulus emulations.
We observe that, in practice, the ZKP modulus is frequently significantly larger than the FHE modulus, especially for smaller circuits.
Because modular reduction produces the same result whether it is applied to the inputs or the outputs of an operation, 
we can wait until just before the results could overflow and only compute and prove the modulo reduction then.
Specifically, we can perform $\floor{\frac{\log_2 p}{\log_2 q}}$ multiplication operations in sequence before needing to reduce.
For small circuits with $\log_2 q \approx 60$ bits and a standard field-based ZKP with $\log_2 p \approx 254$ bits, this enables a 4$\times$ reduction in the number of modulo operations.

Without our optimization, there is little difference between the RNS and non-RNS approaches with respect to the effort required to prove modulus gates.
This is because the cost of proving a modular reduction is roughly linear in the bit-width of the modulus.
However, with our optimization, using the RNS approach allows us to reduce the number of modular reductions even further.
Considering an FHE circuit with $k$ arithmetic operations,
a non-RNS, non-optimized implementation requires $k$ modular reductions of size $\log_2 q$.
Our optimization reduces this to roughly $k \frac{\log_2 q}{\log_2 p}$ modular reductions of size $\log_2 q$.
By RNS-splitting each element into $L$ limbs, the size of each gate is reduced to $\log_2 q_i \approx \frac{\log_2 q}{L}$ but, without any optimizations, the number of modular reductions increases to $kL$, negating the benefits.
However, with our optimization, we require only $kL\frac{\log_2 q_i}{\log_2 p} \approx k\frac{\log_2 q}{\log_2 p}$ modular reductions of reduced size $\frac{\log_2 q}{L}$. 
Therefore, with our optimization, RNS allows us to reduce the cost per modular reduction even while already reducing their number.
For example, with $\log_2 q \approx 60$ bits and $\log_2 p \approx 254$ bits as above, an RNS approach splitting $q$ into two 30-bit moduli would halve the cost again, giving us a total 8$\times$ decrease in modular reduction overhead.
As a result, we can construct comparatively efficient ZKP circuits for a wide range of FHE applications while allowing freedom in FHE and ZKP parameter selection.

\fakeparagraph{Arithmetization}
Given a computation defined as a circuit, different ZKP systems follow different approaches in translating the correctness of computation into an arithmetic proof system.
Since the circuits arising in FHE integrity follow similar patterns, we can generically select an appropriate approach independent of the concrete FHE application.
We propose to use R1CS~\cite{Ben-Sasson2013-oe}, one of the most widespread arithmetization approaches, which converts circuits into a system of Rank-1 constraints.
    Basic arithmetic operations such as additions and multiplications can be realized directly using a single constraint.
    However, more complex operations (e.g., rounding) require a larger number of constraints.
    Since the majority of operations in (NTT-based) FHE are simple arithmetic operations, they can be efficiently translated to R1CS.
    While PLONK-like systems~\cite{Gabizon2022-ba} extend R1CS with support for custom gates that can model commonly recurring sub-circuits, this is only efficient for low-degree sub-circuits\footnote{Very recent work~\cite{Chen2022-gw} promises to address this issue by improving efficiency for higher-degree sub-circuits. However, only an incomplete closed-source proof-of-concept implementation exists at this point.}
    In our setting, the only candidate for such extraction would be the modulo sub-circuit, but this is not a low-degree circuit.
    Beyond R1CS, STARKs~\cite{Ben-Sasson2018-vo} present an alternative way to express computations not as circuits but as a series of data manipulations.
    The efficiency of STARKs is directly related to the size of the state space and the complexity of the transition function describing each step.
    Since FHE features large ciphertext expansion and a high-degree modulo function, this makes it a poor candidate for realization using STARKs.
    Recent work~\cite{Xie2019-ft} can also forgo explicit arithmetization and instead directly express arithmetic circuits. However, this approach scales poorly with the number of inputs, which is high for FHE circuits due to the expansion from individual ciphertexts to $2N$ field elements (where $N\geq2^{13}$).
    As a result, R1CS currently remains the most appropriate choice for FHE integrity applications, and we consider only ZKP systems with efficient support for R1CS arithmetizations.

\fakeparagraph{ZKP Scheme Selection}
In recent years, a large variety of efficient ZKP systems has been proposed, with many seeing widespread use in real-world deployments.
However, when selecting suitable candidates for our instantiation, we need to consider that FHE has slightly different requirements than, e.g., blockchain applications.
For example, due to the large ciphertext expansion, proof size is less of a concern in FHE, which already introduces a noticeable communication overhead.
Instead, we are mostly concerned with achieving a good trade-off between prover and verifier time.
Note that we do not require public verifiability and can therefore exploit potentially more efficient designated verifiability schemes.
We select four candidate approaches that are especially suitable for verifiable FHE:

First, we consider Groth16~\cite{Groth2016-cu}, which represented a major breakthrough in practical ZKP research, showing that zkSNARKs can be achieved with good concrete efficiency.
It requires a trusted setup, which presents difficulties in settings where the set of parties is not known in advance, such as in blockchain applications.
However, in verifiable FHE, the client and server can easily realize this trusted party during a one-time setup via a 2-party maliciously-secure MPC protocol.
Follow-up work has introduced new schemes with different tradeoffs,
such as removing the need to re-run the setup for each circuit.
However, this is not relevant to FHE, where parameters are usually already circuit-specific.

On the other hand, we do evaluate transparent SNARKs that completely remove the need for such a set-up.
Here, we consider Bulletproofs~\cite{Bunz2018-mi} which is part of a generation of zkSNARKS without trusted setup that brings them into the same realm of performance as traditional efficient constructions that require (universal or per-circuit) setup.
While most implementations of Bulletproofs focus on using it for efficient range proofs, for FHE we also require the ability to support more generic R1CS systems.
We also evaluate Aurora~\cite{Ben-Sasson2019-qb}, which is part of the same generation and trades off asymptotically worse prover times and proof sizes for a move to post-quantum secure assumptions, which matches nicely with the post-quantum security of FHE schemes.

Finally, we also consider Rinocchio~\cite{Ganesh2021-rq} by Ganesh et al.,
which we briefly discussed in \Cref{sub:paradigms}.
Rinocchio offers native support for FHE-friendly rings, potentially giving significant performance benefits over systems that need to emulate these rings.
We extend Rinocchio with a more optimized encoding scheme but, for brevity, refer to \Cref{app:encodings} for further details on our optimization.
However, its expressiveness is limited, as Rinocchio only supports arithmetic ring operations, whereas some FHE operations (e.g., relinearization) use component-wise rounding operations internally. 
Additionally, Rinocchio only provides only around 60 bits of (computational) soundness for the rings used in FHE. 
We use a simple soundness amplification strategy, running three separate instances of the protocol to achieve stronger soundness guarantees.
Overall, Rinocchio is much more FHE-friendly than previous proof or argument systems, but still struggles to efficiently represent state-of-the-art FHE optimizations natively.
Nevertheless, we include it because it represents an interesting avenue for future work and promises significantly improved performance for the circuits it can support.
In \Cref{sec:performance}, we evaluate our construction when instantiated using these ZKP schemes and also compare it against a TEE-based instantiation. 

\vspace{-0.3em} %

\begin{table*}[tpb]
\vspace{-1em}
    \centering
    \newcommand{\tableindent}{\hspace{0.8em}}
    \begin{tabular}{rrrrrrrrrr}
        \toprule
        \multicolumn{1}{c}{\multirow{2}{*}{ZKP System}} & \multicolumn{3}{c}{\texttt{Toy}} &   \multicolumn{3}{c}{\texttt{Small}}  &   \multicolumn{3}{c}{\texttt{Medium}}  \\
        & Setup & Prover & Verifier &  Setup & Prover & Verifier &  Setup & Prover & Verifier \\
        \midrule
        FHE [No Integrity] & 0.003~s & 0.002~s & 0.001~s &  0.807~s & 0.011~s & 0.009~s & 1.053~s & 0.014~s & 0.010~s  \\
        Bulletproofs~\cite{Bunz2018-mi}& -\footnotemark[6] & 7569.799~s & 552.079~s & - & 3957.122~s & 278.433~s & - & 8697.741~s & 575.792~s  \\
        Aurora~\cite{Ben-Sasson2019-qb}& - & 1554.589~s & 32.880~s & - & 3750.477~s & 79.323~s & - & 5028.085~s & 106.345~s  \\
        Groth16~\cite{Groth2016-cu} & 198.640~s & 195.941~s & 0.002~s & 479.222~s & 472.711~s & 0.002~s & 642.470~s & 633.741~s & 0.002~s  \\             
        Rinocchio~\cite{Ganesh2021-rq}& 0.485~s & 0.320~s & 0.096~s & 46.700~s & 305.000~s & 0.153~s & 56.90~s & 443.000~s & 0.181~s  \\
        TEE~[Appendix \ref{app:fhe-in-tee-opt}]& - & 0.154~s & - & - & 1.100~s & - & - & 1.260~s & -  \\
        \bottomrule
    \end{tabular}
    \vspace{0.75\baselineskip}
    \caption{Performance results for different instantiations of verifiable Fully Homomorphic Encryption. \\
    For FHE, Setup = Key Generation, Prover = Homomorphic Computation and Verifier = Encryption/Decryption}
    \label{table:performance}
    \vspace{-2.5em}
\end{table*}

\vspace{-0.25em}
\subsection{Verifiable FHE via TEE}
\vspace{-0.5em}
In this instantiation, we use hardware attestation rather than cryptographic proofs to provide the integrity component.
We note that, while there has been a plethora of attacks on TEE-based \emph{confidentiality} guarantees~\cite{Fei2021-xb, Nilsson2020-mg, Murdock2020-dl},
there have been significantly fewer issues with the attestation-based integrity guarantees~\cite{Murdock2020-dl}. %
Natarajan et al. presented the first implementation of this FHE-in-TEE approach~\cite{Natarajan2021-me} and we extend their work to our notion with server inputs.
In addition, we introduce an optimization that allows us to efficiently compute subfunctions on untrusted hardware.
Since TEEs can directly attest to the program that they are running, there is no need for explicit arithmetization. 
However, programs will frequently require adjustments to properly interface with the enclave SDK and to remove unsupported operations and performance bottlenecks.
Nevertheless, TEEs support most FHE schemes more naturally than ZKP systems, including offering native support for, e.g., rounding or floating-point operations.
However, computations and especially memory operations inside the enclave are significantly slower than operations in the untrusted domain.
We address this issue partially by introducing a new optimization that accelerates FHE-in-TEE by a factor of two for batched multiplications.
The key insight in our optimization is that the server can also rely on the guarantees of the enclave to not leak its inputs to the client.
Specifically, it can use lightweight cryptographic proofs that do not have the zero-knowledge property to prove to the enclave that computations performed on the untrusted hardware are indeed correct and can be relied upon by the enclave.
This enables our optimization which computes computationally expensive batched multiplications on untrusted hardware and then uses an efficient Schwartz-Zippel-based proof of correctness to move them back to the trusted domain. For brevity, we refer to \Cref{app:fhe-in-tee-opt} for a detailed description.

\vspace{-0.5em}
\subsection{Performance Analysis}
\vspace{-0.5em}
\label{sec:performance}
In this section, we evaluate our ZKP- and TEE-based instantiations across a range of workloads representing different levels of complexity. 
We conclude our analysis with a brief outlook on the future of FHE integrity.

\footnotetext[6]{We omit setup results when no additional setup is required.}
\stepcounter{footnote}

\fakeparagraph{Implementation \& Setup}
We use the Microsoft SEAL~\cite{Chen2017-xv} implementation of the BGV scheme~\cite{Brakerski2014-qh}, which is a state-of-the-art RNS- and NTT-based implementation.
We express our ZKP circuits using Circom~\cite{Garcia_Navarro2020-jg} which translates its custom specification language to R1CS.
We rely on a variety of state-of-the-art ZKP implementations for the backends:
we use the arkworks library~\cite{arkworks} to implement Groth16~\cite{Groth2016-cu}, for Aurora~\cite{Ben-Sasson2019-qb} we use the libiop library by the same authors, and for Bulletproofs~\cite{Bunz2018-mi} we use the Dalek library~\cite{daklek-bulletproof-lib}.
We implement Rinocchio~\cite{Ganesh2021-rq} and extend it with an optimized encoding scheme (cf. \Cref{app:encodings}). We make our implementation available as open-source\footnote{\url{https://github.com/MarbleHE/ringSNARK}}.
For the TEE-based approach, we re-implement CHEX-MIX~\cite{Natarajan2021-me} and extend it with our optimization (cf. \Cref{app:fhe-in-tee-opt}), targeting Intel SGX via the OpenEnclave SDK~\cite{oe}. 
We make our FHE-in-TEE framework available as open-source\footnote{\url{https://github.com/MarbleHE/FHE-in-TEE}}.
We evaluate our implementations on an AWS \texttt{c5d.4xlarge} instance with 16~vCPUs and 32~GB of RAM.
We make our instantiations and evaluation setup publicly available\footnote{\url{https://github.com/MarbleHE/ZKP-FHE}}.

\fakeparagraph{Workloads}
We consider three different circuits, each representing a different level of complexity:
\emph{(i)}~Our \texttt{Toy} circuit computes a ciphertext-ciphertext multiplication (tensoring only, no post-processing) on two inputs provided by the client, i.e., in the outsourced computation setting considered by previous work. 
\emph{(ii)}~The \texttt{Small} circuit represents a more realistic low-depth two-party computation, computing $\pcalgostyle{NoiseFlood}(x\cdot{}v + w)$ for an encrypted client input $x$ and private server inputs $v$ and $w$ (here, both arithmetic operations are ciphertext-plaintext operations). 
The presence of the server inputs requires input checks to ensure  validity and prevent key-recovery attacks. 
Their private nature requires performing and proving noise-flooding as part of the computation.
We follow the approach described in \cite{Bois2021-qa}, which adds randomly-selected encryptions of zero to increase the noise of the ciphertext without modifying the message.
\emph{(iii)}~Finally, our \texttt{Medium} circuit introduces ciphertext-maintenance operations, computing $\pcalgostyle{NoiseFlood}\left(\pcalgostyle{ModSwitch}\left((x-w)^2\right)\right)$ for a client input $x$ and a private server input $w$. 
As in the previous task, this requires proving the validity of the client input and ensuring server privacy via noise-flooding.
In addition, it requires computing and proving the modulus switching ciphertext operation, going well beyond what prior work on FHE integrity is able to express.

\fakeparagraph{Performance}
In \Cref{table:performance} we present the performance results for the different instantiations, compared against a baseline of non-verified FHE. 
The FHE parameters for the workload are $N=8192$ and $\log_2 q = 137 = 45 + 46 + 46$, and the zero-knowledge proofs have between $2^{22}$ and $2^{24}$ R1CS constraints.
We observe that the runtimes fall into three different categories:
practical (seconds), acceptable (minutes), and impractical (hours).
We note that verifier times are always either practical or at least acceptable.
However, prover time varies wildly between the different instantiations.
For example, transparent SNARKs (Bulletproofs and Aurora) take several hours to compute the proofs. In addition, they have the slowest verifier times, making them overall unattractive for FHE verification, especially when considering that FHE permits a straightforward realization of trusted setups.
Groth16 offers the best verification time, being nearly indistinguishable from pure FHE. 
At the same time, it offers acceptable prover runtimes in the order of minutes.
This comes at the cost of several minutes of trusted setup, but that can be amortized over many client queries.
Rinocchio (implemented with our optimizations) offers slightly slower but still very practical client verification but improves significantly on prover time and especially setup times.
Finally, FHE-in-TEE 
offers by far the most practical performance at the cost of additional trust assumptions and hardware requirements.

\vspace{-0.25em}
\fakeparagraph{Outlook}
Our work shows that, while the cost of robustness is clearly non-negligible, it is, today, already acceptable for high-value applications in challenging settings.
FHE applications currently mostly focus on settings where latency is not critical, which allows them to tolerate the additional overhead more easily.
While our constructions and instantiations explored (and optimized) existing state-of-the-art FHE and ZKP schemes, there are further opportunities to improve performance through the co-design of FHE and ZKP schemes.
More fundamentally, we need to improve our understanding of how to construct efficient and expressive ring-native ZKP systems.
We believe that the increasing demand to deploy FHE in challenging real-world environments will also drive future research on these issues.

\vspace{-0.25em}
\fakeparagraph{Acknowledgments}
We would like to thank Chaya Ganesh, Anca Nitulescu, Eduardo Soria-Vazquez, Michael Steiner, Nojan Sheybani, Erin Hales, Lea Nürnberger, Martha Norberg Hovd, and the PPS Lab team for their insightful input and feedback.
We would also like to acknowledge our sponsors for their generous support, including Meta, Google, SNSF through an Ambizione Grant No. 186050, and the Semiconductor Research Corporation.

\bibliography{refs, refs_add}
\appendices
\crefalias{section}{appendix}

\newpage

\section{} %
\label{app:epscorrectness}
\label{app:furtherdef}
\noindent
In this appendix, we provide additional formal definitions.

\vspace{0.75em}
\begin{definition}[Fully Homomorphic Encryption]
\label{def:fhe}
A \acf{FHE} scheme is a tuple of PPT algorithms $(\kgen, \enc,\dec, \eval)$.
 \begin{itemize}
     \item $\kgen(\secparam) \to (\pk,\sk)$
     \item $\enc_\anykey(x) \to c_x$ where \anykey{} is either \pk{} or \sk{}.
     \item $\eval_\pk(f,c_w, w) \to c_y$ where $y = f(x,w)$ and \\ $w$ are (optional) plaintext inputs.
     \item $\dec_\sk(c_y) \to y$
\end{itemize}
FHE schemes must satisfy \emph{correctness}, \emph{security}, and \emph{compactness} properties.
We omit a formal treatment here and
instead refer to \cite{Gentry2009-zu} for a definition of these properties.
\end{definition}

\vspace{0.75em}
\begin{definition}[FHE Circuit Privacy~\cite{Gentry2009-zu}]
\label{def:fhe-circuit-privacy}
An FHE scheme $\mathcal{E}$ is \emph{circuit-private} for circuits in $\mathcal{F}_\mathcal{E}$ if, for any key-pair (\pk, \sk) output by $\mathcal{E}.\kgen(\secparam)$ any circuit $f \in \mathcal{F}_\mathcal{E}$, and any fixed inputs $c_x$  in the image of $\mathcal{E}.\enc{}$ for inputs $x$, the following are (statistically) indistinguishable:
\[\mathcal{E}.\enc_\pk(f(x)) \approx \mathcal{E}.\eval_\pk(f,c_x)\]
\end{definition}

\vspace{0.75em}
 \begin{definition}[SNARK]
 \label{def:snark}
 Let $\rel$ be an efficiently computable binary relation which consists of pairs of the form $(x, w)$, where $x$ is a statement, and $w$ is a witness. Let $L$ be the language associated with the relation $\rel$, i.e., $L = \setst{ x }{\exists w . \rel(x,w) = 1}$.

\noindent
 A triple of polynomial time algorithms $\Pi = (\setup,\prove, \verify)$ is a SNARK for an NP relation $\rel$, if the following properties are satisfied:

\noindent
Completeness: For every true statement for the relation $\rel$, an honest prover with a valid witness always convinces the verifier:
 $  \forall (x, w) \in \rel{} :$
 \begin{equation*}
 \condprob{
 \verify_\vk(x, \pi) = 1
 }{
 \begin{gathered}
 (\crs, \vk) \gets \setup(\secparam)\\
 \pi \gets \prove_\crs(x, w)
 \end{gathered}
 } = 1
 \end{equation*}
 
\noindent
 Knowledge Soundness: For every PPT adversary, there exists a PPT extractor that gets full access to the adversary's state (including its random coins and inputs). Whenever the adversary produces a valid argument, the extractor can compute a witness with high probability: 
  $\forall \adv{} \exists \mathcal{E} : $
 \begin{equation*} 
 \condprob{
 \begin{gathered}
 \verify_\vk(\tilde{x}, \tilde{\pi}) = 1\\
 \land \rel(\tilde{x}, w') = 0
 \end{gathered}
 }{
 \begin{gathered}
 (\crs, \vk) \gets \setup(\secparam)\\
  ((\tilde{x}, \tilde{\pi}); w') \gets \adv{}|\mathcal{E}(\crs) \\
 \end{gathered}
 } = \negl
 \end{equation*}

\noindent
 We stress here that this definition requires a \emph{non-black-box} extractor, i.e., the extractor gets full access to the adversary's state.

\noindent
Succinctness: For any $x$ and $w$, the length of the proof is given by $|\pi| = \poly \cdot \pcpolynomialstyle{polylog}(|x| + |w|)$. 
 \end{definition}

\vspace{0.75em}
 \begin{definition}[zk-SNARK]
 \label{def:zksnark}
 A zk-SNARK for a relation \rel{} is a SNARK for \rel{} with the following additional property: 

\noindent
Zero-Knowledge: There exists a PPT simulator $\sdv = (\sdv_1, \sdv_2)$ such that $\sdv_1$ outputs a simulated CRS \crs{} and a trapdoor \td{}; On input \crs{}, $x$, and \td{}, $\sdv_2$ outputs a simulated proof $\pi$, and for all PPT adversaries $\adv = (\adv_1, \adv_2)$, such that
 \begin{align*}
 &\left|\condprob{
 \begin{gathered}
 (x, w) \in \rel \\
 {}\land{} \\
 \adv_2(\pi) = 1
 \end{gathered}    
 }{
 \begin{gathered}
 (\crs, \vk) \gets \setup(\secparam) \\
 (x, w) \gets \adv_1(\secparam, \crs) \\
 \pi \gets \prove_\crs(x, w)
 \end{gathered}
 }
 - \right.
 \\
 &\left.\condprob{
 \begin{gathered}
 (x, w) \in \rel \\
 {}\land{} \\
 \adv_2(\pi) = 1
 \end{gathered}
 }{
 \begin{gathered}
 (\crs', \td) \gets \sdv_1(\secparam) \\
 (x, w) \gets \adv_1(\secparam, \crs') \\
 \pi \gets \sdv_2(\crs', \td, x)
 \end{gathered}
 }
 \right| = \negl
 \end{align*}
 \end{definition}

\vspace{0.75em}
\begin{definition}[$\epsilon$-Correctness]
\label{def:eps-correctness}
A scheme is correct if any honest computation will decrypt to the expected result.
For approximate schemes, there is a slight error $\epsilon$, this is zero for standard exact schemes.

More formally, a scheme is \emph{correct} if for all functions $f$, and for all $x$ in the domain of $f$: 
\begin{equation*}
\condprob{
\begin{gathered}
\norm{\dec_\sk(c_y) - f(x)} \le \varepsilon
\end{gathered}
}{
\begin{gathered}\
(\pk, \sk) \gets \kgen(\secparam, f)\\
(c_x, \tau_x) \gets \enc_\anykey(x) \\
(c_y, \tau_y) \gets \eval_\anykey(c_x)
\end{gathered}
} = 1
\end{equation*} 

where $\norm{\cdot}$ is a scheme-specific norm, and $\varepsilon$ is a scheme-specific upper bound on the decoding error (which may depend on $f$, $\pk$, or other quantities of the scheme). 
\end{definition}

\vspace{0.75em}
\begin{definition}[vFHE with Server Circuit Privacy]
A \emph{maliciously-secure verifiable FHE scheme with server circuit privacy} is a tuple $(\kgen, \enc, \verify, \dec, \eval)$ of PPT algorithms:
\begin{itemize}
\item $\kgen(\secparam, \mathcal{F}) \to (\pk, \sk)$ 
\item \enc, \dec, and \verify{} are defined as in \Cref{def:vfhe}
\item $\eval_{\pk}(c_x, f) \to (c_y, \tau_y)$ where  $y = f(x)$ for $f \in \mathcal{F}$
\end{itemize}

\noindent
The scheme must satisfy the 
\emph{correctness}, %
\emph{completeness}, %
\emph{soundness}, %
and \emph{security} %
properties defined in \Cref{sec:notions}.
In addition, it must satisfy the \emph{server circuit privacy} property defined below.
\end{definition}

\vspace{0.75em}
\begin{definition}[Server Circuit Privacy]
    Informally, a scheme offers circuit privacy when \eval{} reveals nothing about the function.
    We assume a semi-honest PPT adversary with access to all keys, including \sk.
    Formally, a scheme offers \emph{server circuit privacy} when, for all $f \in \mathcal{F}$, the following are (statistically) indistinguishable:
    \[ \eval_\pk(c_x,f) \approx \eval_\pk(c_x,f') \]  
\end{definition}

\section{} %
\label{app:generic-constructions}

\noindent
In this appendix, we provide generic constructions for the remaining variants of our verifiable FHE notion. We refer to the extended version of this paper for formal proofs~\cite{extended-version}.
We begin with a formal description of the construction for our core notion.
As described in \Cref{sec:generic-constructions}, the ZKP system could also be replaced with homomorphic MACs due to the lack of server inputs.
\stepcounter{construction}
\begin{construction}[vFHE from ZKP]
\begin{itemize}
    \item Let \fhe{} = (\fhe.\kgen, \fhe.\enc, \fhe.\dec, \fhe.\eval) be an \\ \indcpa{} secure FHE scheme.
    \item Let \zkp{} = (\zkp.\kgen, \zkp.\prove, \zkp.\verify) be a proof of knowledge (e.g., a SNARK) for the relation \\
    $\left\{ \left((f, c_x, c_y), w\right) : c_y = \fhe.\eval_{\pk_\fhe}(f, c_x, w) \land w \in \mathcal{W} \right\}$
\end{itemize}

\noindent
For  $f \in \mathcal{F}$, we construct a vFHE scheme $(\kgen, \enc, \eval, \\\verify, \dec)$ satisfying \Cref{def:vfhe} from these building blocks as follows:
\begin{itemize}
\item $\kgen(\secparam, f) \to \left((\pk_\fhe, \pk_\zkp), (\sk_\fhe, \sk_\zkp)\right)$, where \\
$(\pk_\fhe, \sk_\fhe) = \fhe.\kgen(\secparam)$ and\\
$(\pk_\zkp, \sk_\zkp) = \zkp.\kgen(\secparam, f)$
\item $\enc_\pk(x) \to \left(c_x, \tau_x\right)$, where $\tau_x = c_x$ and $c_x = \left(\fhe.\enc_{\pk_\fhe}(\pt)\right)_{\pt \in x}$ for $x \in \mathcal{X}$
\item $\eval_\pk(c_x, w) \to (c_y, \tau_y)$, where $c_y = \fhe.\eval(f,c_x,w)$ and $\tau_y = \zkp.\prove(c_y, \tau_x,w)$
\item $\verify_\sk(c_y, \tau_x, \tau_y) = \zkp.{\verify_\zkp.\sk}\left((\tau_y, \tau_x)\right)$. 
\item $\dec_\sk(c_y) \to y$, where $y = \fhe.\dec_{\sk_\fhe}(c_y)$
\end{itemize}
\end{construction}

\noindent
We note that we can construct a scheme with $\epsilon$-correctness following the same approach if we require $\mathcal{E}$ to be an approximate FHE scheme with \indcpad{} security~\cite{Li2022-gc}.\\

\noindent
Our constructions for vFHE with (client or server) circuit privacy are directly derived from the construction for the core notion but instantiated with a universal function that takes the (private) function as part of the input. %

\begin{construction}[vFHE with Client Circuit Privacy]
We instantiate a vFHE scheme according to \Cref{def:vfhe} with $f = F_U$, a universal circuit (for a class of functions $\mathcal{F}$).
In order to compute $f(x)$ for $f \in \mathcal{F}$ we  use the input $c'_x = \enc_\anykey(x || f)$.
\end{construction}

\begin{construction}[vFHE with Server Circuit Privacy]
We instantiate a vFHE scheme with server privacy according to \Cref{def:vfhesp} with $f = F_U$, a universal circuit (for a class of functions $\mathcal{F}$).
In order to compute $f(x, w)$ for $f \in \mathcal{F}$ we  use the server input $w' = w || f$.
\end{construction}

\noindent
Since we use a generic ZKP system to enforce validity checks on the server inputs, extending our construction to verifiable FHE with input predicates is straight-forward:

\vspace{0.75em}
\begin{construction}[vFHE with Input Predicates]
\begin{itemize}
    \item Let \fhe{} = (\fhe.\kgen, \fhe.\enc, \fhe.\dec, \fhe.\eval) be an \\ \indcpa{} secure FHE scheme with (FHE) circuit privacy (See \Cref{app:furtherdef} for formal definitions).
    \item Let $\mathcal{F_{\mathcal{X},\mathcal{W}}}$ be a set of functions, with  $f \in \mathcal{F}$ if\footnote{Note that inverse does not have to hold. A practical realization could rely on one-sided bounds on noise growth.}
    for all $x \in \mathcal{X}$ and $w \in \mathcal{W}$, $\fhe.\dec_{\fhe.\sk}(\fhe.\eval(c_x, w))$ results in a valid plaintext (which can be a subset of all possible plaintext). 
    \item Let $\phi{}$ be a predicate on the inputs.
    \item Let \zkp{} = (\zkp.\kgen, \zkp.\prove, \zkp.\verify) be a generic  ZKP with ZKP$\left\{ w : \fhe.\eval_{\pk_\fhe}(c_x,w) \
    \land \ w \in \mathcal{W}  \land \phi(w) \right\}$.
\end{itemize} 

\noindent
For  $f \in \mathcal{F}$ , we construct a vFHE scheme with input predicates $(\kgen, \enc, \eval, \verify, \dec)$ satisfying \Cref{def:vfheip} from these building blocks as follows:
\begin{itemize}
\item $\kgen(\secparam, f) \to \left((\pk_\fhe, \pk_\zkp), (\sk_\fhe, \sk_\zkp)\right)$, where \\
$(\pk_\fhe, \sk_\fhe) = \fhe.\kgen(\secparam, f)$ and\\
$(\pk_\zkp, \sk_\zkp) = \zkp.\kgen(\secparam, f)$
\item $\enc_\pk(x) \to \left(c_x, \tau_x\right)$, where $\tau_x = c_x$ and $c_x = \left(\fhe.\enc_{\pk_\fhe}(\pt)\right)_{\pt \in x}$ for $x \in \mathcal{X}$
\item $\eval_\pk(c_x, w) \to (c_y, \tau_y)$, where $c_y = \fhe.\eval(f,c_x,w)$ and $\tau_y = \zkp.\prove(c_y, \tau_x,w)$
\item $\verify_\sk(c_y, \tau_x, \tau_y) = \zkp.{\verify_\zkp.\sk}\left((\tau_y, \tau_x)\right)$. 
\item $\dec_\sk(c_y) \to y$, where $y = \fhe.\dec_{\sk_\fhe}(c_y)$
\end{itemize}
\end{construction}

\section{} %
\label{app:optimizations}
\label{sub:rinocchio-batched-encoding}
\label{app:encodings}
\noindent
In the following, we describe our optimizations for the Rinocchio protocol by Ganesh et al.~\cite{Ganesh2021-rq}.
The original paper introduces two possible encodings for the cyclotomic rings used by FHE. 
The first one (dubbed ``Regev-style'' encoding) encodes each of the $N$ coefficients in $\ZZ_q$ by encrypting it into an element of $\ZZ^n_Q$ using a LWE cryptosystem scheme; the parameters of the encoding scheme are chosen to ensure that the encodings are $k$-linearly-homomorphic, where $k$ is determined by the circuit.  
The second construction (``Torus encoding'') uses a variant of the TFHE cryptosystem. 

The Regev encoding has an expansion factor of $\frac{N\cdot n \cdot \log_2(Q)}{N \cdot \log_2(q)} = n \cdot \log_q(Q)$, as it encodes each of the $N$ coefficients in $\ZZ_q$ as an element of $\ZZ_Q^n$. 
However, most FHE implementations will not be able to support a plaintext modulus of the size of $q$ (typically hundreds of bits), and in practice one would need to encode each of the $l$ CRT components individually, leading to an expansion factor of $ l \cdot n \cdot \log_q(Q)$. 
Using this encoding will thus slow down the prover and verifier significantly, as all encodings, decodings, and computations over the encoding space will be slow. 

The TFHE encoding, on the other hand, requires using floating-point arithmetic to encode and decode, unlike the FHE scheme used for computation. 
Additionally, this encoding does not allow us to use the (potentially heavily optimized) ring arithmetic libraries provided by the FHE library. 

Therefore, we propose a new RLWE Regev-style encoding for Rinocchio, taking advantage of the batching technique commonly used in FHE. 
For many FHE schemes, if the plaintext modulus $t$  satisfies the condition $t = 1 \mod 2N$, one can use an efficient encryption that packs $N$ plaintext values (interpreted as an element of $R_t$) into a single ciphertext in $R_q^2$. 
For our encoding, we take an input in $R_q$ as $l$ polynomials in $R_{q_1}, \ldots, R_{q_l}$ (this decomposition is already used natively by the FHE scheme for efficiency reasons), and encode each of those polynomials as an element in $R_Q$. 
The expansion factor in this case is $\frac{l \cdot \log_2(Q)}{\log_2(q)} = l \cdot \log_q(Q)$, improving on the Regev encoding by a factor of $N \ge 2^{10}$. 
Using this batching technique imposes the requirement $q_i = 1 \mod 2N$ on the ciphertext moduli of the FHE scheme; this condition is already necessary for some schemes (e.g., RNS-optimized BGV \cite{Kim2021-mm}), and can be easily satisfied for all other schemes.

\section{} %
\label{app:fhe-in-tee-opt}
\noindent
In the following, we describe our optimization for FHE-in-TEE:
Executing any code inside a TEE incurs a slowdown (due to reduced computational power and memory), especially in the case of FHE computations, that are typically compute- and memory-intensive. 
To alleviate this slowdown, we propose a new method to accelerate FHE computations inside TEEs by taking advantage of faster (but untrusted) hardware (e.g., a vanilla untrusted CPU, a CPU with specialized vector instructions repurposed for FHE \cite{hexl}, a GPU accelerator for FHE \cite{Ozerk2021-pl}, or even a dedicated hardware accelerator). 

The key insight to our improvement is that both the TEE and the untrusted hardware are on the side of the (malicious) server. 
Therefore, the server's input does not need to be protected from the server, and can be stored on the server's own untrusted hardware; the client's inputs are only available in their encrypted form, and can thus also be stored outside the enclave. 
This insight allows us to devise a protocol for \emph{verifiably} outsourcing certain FHE operations. 
In order to do this efficiently, we rely on a very lightweight, information-theoretic argument of equality, based on the generalized Schwartz–Zippel lemma over rings: 
\begin{theorem}[Generalized SZ over Rings~\cite{Bishnoi2018-co,Ganesh2021-rq}]\label{thm:schwartz-zippel}
For a ring $R$, let $f: R^n\to R$ be a $n$-variate non-zero polynomial, let $A \subseteq R$ be a finite exceptional set, and let $\deg(f)$ denote the total degree of $f$. 
Then: 

$$
\probsub{\vec{a}\gets A^n}{f(\vec{a}) = 0} \le \dfrac{\operatorname{deg}(f)}{|A|}
$$
\end{theorem}

For a given computation, we encode the expected result in one polynomial ($f$), and the actual result computed on untrusted hardware in another polynomial ($g$). 
The trick for efficiency, then, is to compute the compute and compare $f(a)$ and $g(a)$ faster than computing the full representation of $g$ in the first place. 

\balance
Consider, for example, the \emph{tensoring} operation, which is the most computationally expensive part of FHE multiplication (for the B/FV, BGV, and CKKS schemes). 
In the following, we will interpret a ciphertext $\ct = (\ct_0, \ldots, \ct_{k-1}) \in R_q^k$ as a polynomial of degree $k-1$ over $R_q$, where $\ct_i$ is the $i$-th coefficient. 

The tensoring operation takes as input two ciphertexts $\ct = (\ct_0, \ct_1), \ct' = (\ct'_0, \ct'_1) \in R_q^2$, and outputs $\cout = \ct \cdot \ct' = (\ct_0 \cdot \ct'_0, \ct_0 \cdot \ct'_1  + \ct'_0 \cdot \ct_1, \ct_1 \cdot \ct'_1) \in R_q^3$. 
Now, evaluating the expected result $\ct \cdot \ct'$ at a random point $a \in A$ can be done efficiently as $f(a) =  (\ct \cdot \ct')(a) = \ct(a) \cdot \ct'(a) = (\ct_0 + a \cdot \ct_1) \cdot (\ct'_0 + a \cdot \ct'_1)$.
Evaluating the untrusted result $\ct_\text{out}$ at this same point can be done using Horner's rule:
$g(a) := \ct''(a) = \ct''_0 + a \cdot (\ct''_1 + a \cdot \ct''_2)$
After checking that $f(a) = g(a)$, we know that $\ct \cdot \ct' = \ct''$ with high probability (for the FHE schemes discussed in this paper, $|A| = q_1 \approx2^{60}$, i.e., 60 bits of statistical soundness). 
While computing the result has a concrete complexity of $1 {~}_{R}+_{R}$, $4 {~}_{R}\times_{R}$, verifying the result as outlined above only requires $4 {~}_{A}\times_{R}$, $4 {~}_{R}+_{R}$, $1 {~}_{R}\times_{R}$.

This approach can also be extended as follows to verify $k$ tensoring operations at the same time. Let 
$f(a_1, \ldots, a_k) = \sum_{i=1}^k (\ct_i\cdot\ct'_i)(a_i)
= \sum_{i=1}^k (\ct_{i,0} + a_i \cdot \ct_{i,1})\cdot (\ct_{i,0} + a_i \cdot \ct'_{i,1}),$
and define
$g(a_1, \ldots, a_k) = \sum_{i=1}^k \ct''_i(a_i) = \sum_{i=1}^k (\ct''_{i,0} + a_i \cdot (\ct''_{i,1} + a_i \cdot \ct''_{i,2}))$.

Computing $k$ tensoring operations has a concrete complexity of 
$ k {~}_{R}+_{R}, 4k {~}_{R}\times_{R},$
while verifying the result by computing $f(\vec{a})$ and $g(\vec{a})$ has a complexity of 
$ 4k {~}_{A}\times_{R}, (6k -2){~}_{R}+_{R}, k {~}_{R}\times_{R}$.
By trading expensive $R$-$R$ multiplications for cheaper $R$-$R$ additions and $A$-$R$ multiplications, we are able to achieve a non-negligible speed-up, which we quantify in the next section. 

We can view our protocol as a much more efficient, non-zero-knowledge version of Rinocchio; indeed, Rinocchio also uses \Cref{thm:schwartz-zippel}, but requires significantly more protocol machinery in order to achieve zero-knowledge. 
In addition, Rinocchio offers roughly $\log_2 q_1 \approx 60$ bits of computational soundness (and thus requires a soundness amplification strategy), while our protocol offers $\log_2 q_1 $ bits of \emph{statistical soundness}, and can therefore provide a satisfactory level of security by itself. 

We note that this optimization is similar to the Slalom framework by Tramèr and Boneh \cite{Tramer2018-vz}, which offloads matrix multiplications (over unencrypted) values to untrusted hardware by using Freivalds' algorithm. 
Slalom relies on the TEE  both for integrity and data confidentiality and  only supports matrix multiplication, whereas our protocol does not require confidentiality of the data stored on the TEE, and can handle arbitrary polynomial computations.

\end{document}